\documentclass[11pt,a4paper]{article}
\usepackage[noadjust]{cite}

\usepackage[margin=2.8cm,bottom=3.5cm]{geometry}

\usepackage{amsmath, amssymb}
\usepackage{color}
\pdfoutput=1 

\usepackage{graphicx} 
\usepackage{subcaption}
\usepackage[T1]{fontenc}

\title{Resonance mediated by fermions in kink-antikink collisions}

\author{Dionisio Bazeia \\
        Departamento de F\'isica, Universidade Federal da Paraíba,\\
        58051-970 Jo\~ao Pessoa, PB, Brazil\\
        bazeia@fisica.ufpb.br
            \and
        Jo\~ao G. F. Campos \\
        Departamento de F\'isica, Universidade Federal da Paraíba,\\
        58051-970 Jo\~ao Pessoa, PB, Brazil\\
        joaogfc@gmail.com
            \and
        Azadeh Mohammadi\\
        Departamento de F\'isica, Universidade Federal de Pernambuco,\\
        Av. Prof. Moraes Rego, 1235, Recife - PE - 50670-901, Brazil\\
        azadeh.mohammadi@ufpe.br
}

\begin{document} 
\maketitle

\begin{abstract}

We investigate generalizations of the $\phi^4$ and sine-Gordon models, including interactions with Dirac Fermions. We observe new resonance phenomena by taking the fermion back-reaction into account. First, we show that the vibrational mode responsible for the resonance structure of the $\phi^4$ model has the same frequency as the energy of the fermion excited state when the back-reaction becomes more significant. Second, we consider the sine-Gordon model with the addition of a fermion field and find that a resonant structure appears, despite the absence of a scalar vibrational mode. The vibrational frequency of the mode responsible for the exchange mechanism is again the energy of the fermion excited state. Therefore, we find a new type of resonant energy exchange mechanism which is mediated by fermions.

\end{abstract}

\section{Introduction}

In field theories, there are many extended but localized structures that are solutions to the equations of motion. They can be topological, such as kinks, vortices, and baby-skyrmions, and non-topological, such as lumps and Q-Balls \cite{manton2004topological, shnir2018topological}. Moreover, they can exhibit particle-like behavior. In the so-called integrable models, these objects behave elastically when scattered and are called solitons. On the other hand, in the non-integrable models, the collisions are highly inelastic, and complex behaviors are observed.

Perhaps the simplest and most investigated localized structures are kinks. They are objects in one spatial dimension which interpolate between two domains with minimal energy. Experimental realizations of kinks include Josephson Junctions \cite{ustinov1998solitons}, buckled graphene \cite{yamaletdinov2017kinks}, gold dislocations \cite{el1987double}, properties of DNA \cite{yakushevich2006nonlinear}, and many others. Moreover, domain walls in ferromagnets are described by kinks in the direction perpendicular to the wall \cite{kardar2007statistical}.

The sine-Gordon model engenders kinks, the prototypes of solitons with one spatial dimension. On the other hand, the $\phi^4$ model is the prototype of kinks interacting inelastically. Kinks can either separate or annihilate in the latter model after a collision. Surprisingly there are windows in parameter space where separation occurs after more than a single bounce. They are called resonance windows. Such behavior was extensively studied in a triplet of seminal papers by Campbell and collaborators \cite{campbell1983resonance, campbell1986kink, peyrard1983kink}. They gave compelling evidence that the resonance windows occur due to an exchange between the kink's vibrational and translational modes. Moreover, there was much evidence in their analysis that $n+1$-bounce windows occur at the border of $n$-bounce windows, forming a fractal structure. After several years, Manton and collaborators could show that a collective coordinate approach corroborates this picture \cite{manton2021kink, manton2021collective}. Resonant behavior in the kink-antikink collision was later found in a large number of studies. A few examples involve deformed $\phi^4$ models \cite{simas2016suppression, bazeia2018scattering}, a deformed $\phi^6$ model \cite{demirkaya2017kink}, the $\phi^8$ model \cite{gani2015kink}, hyperbolic models \cite{bazeia2019kink, bazeia2020oscillons}, and the double sine-Gordon model \cite{gani2018scattering, gani2019multi, simas2020solitary, campos2021wobbling}.

The picture of resonant behavior described by Campbell and collaborators is not the end of the story. So far, other authors have found a few exceptions to the resonant energy exchange mechanism. Perhaps the most important exception occurs in the $\phi^6$ model \cite{dorey2011kink}. The kinks exhibit resonant behavior in such a model, but a vibrational mode is absent when they are analyzed in isolation. Dorey and collaborators showed that resonance occurs because the kink-antikink pair have vibrational modes that can store energy. Later, the same phenomenon was found in other models with asymmetric kinks \cite{christov2021kink, bazeia2021semi}. Resonance windows also occur during interactions with boundaries and impurities which have vibrational modes \cite{kivshar1991resonant, fei1992resonant, fei1992resonant2, goodman2004interaction, arthur2016breaking, dorey2017boundary, lima2019boundary}

More recently, it was shown that resonance windows might survive if the vibrational mode becomes a quasinormal mode with a small decay rate \cite{dorey2018resonant}. This result was corroborated by a toy model, which allows the quasinormal mode to be evaluated analytically \cite{campos2020quasinormal}. Another exception to the exchange mechanism was found recently in a model with an impurity that preserves the kink's Bogomol'nyi–Prasad–Sommerfield (BPS) property \cite{adam2021sphalerons}. In this work, Adam and collaborators showed that the presence of an unstable static solution, known as sphaleron, can store energy on its vibrational modes and produce resonance windows.

In the present work, we report a new type of energy exchange mechanism. We have found that coupling the kink to a fermion which has a massive bound state also leads to the appearance of resonance windows. Our conclusion is illustrated using both the $\phi^4$ and sine-Gordon scalar fields interacting with fermions.

Systems composed of topological structures and fermions have many interesting properties. The fractional fermion number was shown to appear due to the presence of fermion zero modes in the seminal paper by Jackiw and Rebbi \cite{jackiw1976solitons}. In some cases, these modes are guaranteed to exist by the index theorem \cite{atiyah1963index}. One of the most important experimental realizations of fermion-kink systems appears in polymers, where fractional quantum numbers are observed \cite{niemi1986fermion}. There have been many advances in the analysis of scattering behavior of generalizations of kinks and lumps in systems with multiple scalar fields \cite{halavanau2012resonance, alonso2018reflection, alonso2020non, alonso2021kink}. However, not many works discuss the addition of fermions in kink-antikink collisions. Works considering the fermion back-reaction are even more scarce. A few relevant studies are described below. 

One widely used approximation in the literature is the neglection of the fermion back-reaction to the kink. With this approximation, bound states between fermions and the $\phi^4$ kink can be obtained analytically for the Yukawa interaction. The detailed solution can be found in Refs.~\cite{charmchi2014complete, charmchi2014massive}. In other kink backgrounds, the fermion spectrum can be computed numerically \cite{bazeia2017fermionic, bazeia2021fermions}. In Refs.~\cite{chu2008fermions, brihaye2008remarks}, the authors studied the interaction of fermions with a frozen kink-antikink configuration by applying the same approximation. The dynamical case was considered in Refs.~\cite{gibbons2007fermions, saffin2007particle, campos2020fermion}. Moreover, the interaction between fermions and wobbling kinks was considered in \cite{campos2021fermions}.

When the back-reaction is included, one cannot find the fermion bound states analytically, and numerical methods are required. This was done in Refs.~\cite{amado2017coupled, klimashonok2019fermions, perapechka2020kinks}. The authors have shown that the effect of the back-reaction can be significant and even create bound kink-antikink pairs. Similar studies have been performed for baby Skyrmions interacting with fermions \cite{perapechka2018soliton, perapechka2019fermion}. Recently, the kink-antikink collisions with fermion back-reaction were analyzed for the first time in Ref.~\cite{campos2022kink}. It was shown that the interaction could be strongly altered by the back-reaction as well. Here, we build upon that work and construct similar models where the fermion excited state is present while the kink's vibrational mode has a different frequency or is absent. Hence, we can investigate whether the fermion excited state can mediate the energy exchange mechanism. Especially it becomes more evident in the case of sine-Gordon, which is an integrable model.

Our work is organized as follows. In the next section, we discuss the models we will analyze, which are generalizations of the $\phi^4$ and sine-Gordon models including a Dirac field. In section \ref{sec_res}, we describe the numerical simulations of kink-antikink collisions. Finally in section \ref{sec_conc} we end with some concluding remarks.

\section{Model}

\subsection{Definition and equations of motion}

We are interested in fermion-kink models which can be written generally as follows
\begin{equation}
S=\int d^2y~\left(\frac{1}{2}\partial^\mu\phi\partial_\mu\phi+\frac{1}{2}i\bar{\Psi}\gamma^\mu\partial_\mu\Psi-\frac{g}{2}G\left(\frac{\phi}{\alpha}\right)\bar{\Psi}\Psi-m^2\alpha^2V\left(\frac{\phi}{\alpha}\right)\right).
\end{equation}
It contains a scalar field subject to a potential $V$, a massless Dirac field, and an interaction between the two fields, which depends on the function $G$. The constants $g$, $m$, and $\alpha$ need to be adjusted for the system of interest. As we will consider the effect of the fermion back-reaction to the kink, the normalization of the fermion field should be fixed. We will adopt
\begin{equation}
\int_{-\infty}^{\infty}\Psi^\dagger(y,\tau)\Psi(y,\tau)dy=1.
\end{equation}

To simplify our model and work with dimensionless fields, couplings and coordinates, we rescale them as follows. We set $\phi\to\alpha\chi$ and $y^\mu\to m^{-1}x^\mu$. To maintain the normalization as unit the Dirac field is rescaled as $\Psi\to m^{1/2}\psi$. The coupling between the Dirac and scalar fields is redefined as $g\to mg$. Thus, we end up with the following Lagrangian density
\begin{equation}
{\cal L}=\frac{\alpha^2}{2}\partial^\mu\chi\partial_\mu\chi+\frac{1}{2}i\bar{\psi}\gamma^\mu\partial_\mu\psi-\frac{g}{2}G(\chi)\bar{\psi}\psi-\alpha^2V(\chi),
\end{equation}
supplemented by the normalization condition
\begin{equation}
\int_{-\infty}^{\infty}\psi^\dagger(x,t)\psi(x,t)dx=1.
\end{equation}
The corresponding equations of motion are
\begin{align}
&i\gamma^\mu\partial_\mu\psi-gG(\chi)\psi=0,\\
&\partial_\mu\partial^\mu\chi+V^\prime(\chi)+\frac{g}{2\alpha^2}G^\prime(\chi)\bar{\psi}\psi=0.
\label{eq_eom_chi}
\end{align}
We find that the only effect of the parameter $\alpha$ is to set the strength of the fermion back-reaction, which is represented by the last term in eq.~\eqref{eq_eom_chi}.

\subsection{Kink solution and fermion bound states}

We will adopt the following representation for the Dirac matrices $\gamma^0=\sigma_1$, $\gamma^1=i\sigma_3$. The static solutions containing fermions and a kink can be found using the following ansatz
\begin{equation}
\psi(x,t)=e^{-iEt}\begin{pmatrix}
\psi_+\\
\psi_-
\end{pmatrix},
\end{equation}
where $\psi_\pm$ are real in this representation. We will suppose that the scalar field possesses two or more minima. If we ignore the back-reaction, the scalar field becomes a background field for the fermion. The kink solution without back-reaction will be called $\chi_k(x)$. Let us choose $G$ such that $G(\chi_k(x))=\tanh(ax)$, where $a$ is a constant. As we will see shortly, this applies to both $\phi^4$ and sine-Gordon models with our choice of the interaction forms.
Hence, the Dirac equation reads
\begin{equation}
\label{eq_dirac_analytic}
\begin{pmatrix}
-\partial_x-g\tanh(ax)&E\\
E&\partial_x-g\tanh(ax)
\end{pmatrix}
\begin{pmatrix}
\psi_+(x)\\
\psi_-(x)
\end{pmatrix}=0
\end{equation}
It implies that the field components are solutions of a set of Schr\"{o}dinger-like equations with a modified  P\"{o}schl-Teller potential
\begin{equation}
-\partial_x^2\psi_\pm+\left[g^2-\frac{g\left(g\pm a\right)}{\cosh^2(ax)}\right]\psi_\pm=E^2\psi_\pm.
\label{eq:PTfermion}
\end{equation}
The two equations are decoupled but have the same eigenvalue $E$.
This system can be solved analytically as shown, for instance, in Refs.~\cite{morse1953methods, chu2008fermions, charmchi2014complete}. The discrete eigenvalues are given by
\begin{equation}
E^{(0)}_{\pm n}=\pm\sqrt{2gan-a^2n^2},\quad n=0,1,2,\ldots<\frac{g}{a}.
\end{equation}
The corresponding eigenfunctions will be denoted by $\psi_{\pm n}^{(0)}(x)$. Explicit expressions can be found in Ref.~\cite{charmchi2014complete}.

If we include the back-reaction, the solutions can only be found numerically \cite{amado2017coupled, klimashonok2019fermions}, except for the zero mode, where the back-reaction vanishes. In such a case, the solution for the scalar field is again $\chi_k(x)$, and the fermion zero mode is given by
\begin{equation}
\psi_0(x)=\mathcal{N}\begin{pmatrix}
\cosh(ax)^{-g/a}\\
0
\end{pmatrix},
\end{equation}
where $\mathcal{N}$ is the normalization constant. The other solutions which include the back-reaction will be denoted by $\psi_{\pm n}(x)$ with eigenvalue $E_{\pm n}$ and kink profile $\chi_n(x)$. We will consider kink-antikink collision where the fermion starts at the zero mode.

\subsection{Stability Equation}
\label{sec:stab}
The spectrum of perturbations around static solutions can be obtained by writing the scalar and fermion fields as some solution plus a small perturbation, i.e., $\chi(x,t)=\chi_n(x)+\eta(x,t)$ and $\psi(x,t)=e^{-iE_nt}(\psi_n(x)+\zeta(x,t))$. Substituting them in the equations of motion, we find the following equations
\begin{align}
    \partial_{t}^2\eta&=\partial_{x}^2\eta-V^{\prime\prime}(\chi_n)\eta-\frac{g}{2\alpha^2}G^{\prime\prime}(\chi_n)\bar{\psi}_n\psi_n\eta-\frac{g}{\alpha^2}G^\prime(\chi_n)\text{Re}(\bar{\psi}_n\zeta),\label{bos-pert}\\
    \partial_t\zeta&=iE_n\zeta-\gamma^0\gamma^1\partial_x\zeta-i\gamma^0gG(\chi_n)\zeta-i\gamma^0gG^\prime(\chi_n)\psi_n\eta.
\end{align}
The term with $G^{\prime\prime}$ in eq. (\ref{bos-pert}) is automatically zero for the $\phi^4$ model.

The initial configuration contains fermions at zero mode. This is also the predominant fermion state during the evolution of the system. Therefore, we specialize to $n=0$, which significantly simplifies the equations. Using the ansatze $\eta(x,t)=e^{-i\omega t}\eta(x)$ and $\zeta(x,t)=e^{-i\omega t}\zeta(x)$, we find the system of eigenvalue equations
\begin{align}
\label{eq:boson-eig}
    \omega^2\eta&=-\partial_{x}^2\eta+V^{\prime\prime}(\chi_k)\eta+\frac{g}{\alpha^2}G^\prime(\chi_k)\text{Re}(\bar{\psi}_0\zeta),\\
    \omega\zeta&=-i\sigma_2\partial_x\zeta+\sigma_1gG(\chi_k)\zeta+\sigma_1gG^\prime(\chi_k)\psi_0\eta.\label{eq:fermion-eig}
\end{align}
This system of equations can only be solved numerically. To solve an eigenvalue problem with an operator that takes the real part, we will split $\zeta$ into the real and imaginary parts and analyze the eigenvalue problem for four real fermionic fields instead of two complex ones. Moreover, we will also define $\xi=\dot{\eta}$ to write the bosonic equation as two first-order equations instead of a second-order one. We will apply a numerical method that consists of discretizing space using a five-point stencil approximation for the derivatives and diagonalizing the resulting matrix using the NumPy library in Python. However, care must be taken when interpreting such states because we are diagonalizing an operator that is neither hermitian nor symmetric. Therefore, different modes may overlap and cannot be interpreted as independent. Moreover, notice that the bosonic and fermionic sectors have different threshold values that separate the discrete spectrum and the continuum. Hence, the method may mask fermionic states if the bosonic threshold is too low and vice-versa.

If we ignore for now the $\alpha$-dependent terms (taking the limit $\alpha \to \infty$ for example), the system decouples, and the spectrum can be found analytically for the following reasons. The bosonic field does not depend on $\zeta$ and obeys a Schr\"{o}dinger-like equation with a modified P\"oschl-Teller potential for both sine-Gordon and $\phi^4$ models. The bosonic solutions can be used to construct solutions to the full decoupled problem by adding a particular solution of eq.~\eqref{eq:fermion-eig} with the corresponding values of $\eta$ and $\omega$. For $\omega=0$, the solution is guaranteed to exist due to the translational invariance of the system and can be found in closed form. For the remaining bosonic states, our numerical analysis indicates that it is possible to find fermionic field configurations such that the full decoupled equations are solved, although without closed form. The remaining discrete solutions can be found by setting $\eta=0$. In such a case, the two components of the fermionic field obey eq.~\eqref{eq:PTfermion}, which has analytical solutions. The decoupled solutions relevant in the following analysis are listed in tables \ref{tab:P4-eig} and \ref{tab:SG-eig}.

Now, what is the effect of the $\alpha$-dependent term? Our numerical analysis indicates that the discrete spectrum listed in the tables remains unaltered when such a term is included. Therefore, it is possible to find if a fermionic or bosonic mode mediates the exchange mechanism by comparing the measured exchange frequency with the ones on the tables.

\subsection{$\phi^4$ model}

In the $\phi^4$ model, the potential is given by
\begin{equation}
V(\chi)=\frac{1}{4}(\chi^2-1)^2.
\end{equation}
The kink solution for configurations where the back-reaction vanishes is
\begin{equation}
\chi_k=\tanh\left(\frac{x}{\sqrt{2}}\right).
\end{equation}
Moreover, the coupling that leads to the Dirac equation in \eqref{eq_dirac_analytic} is Yukawa $G(\chi)=\chi$, leading to $a=1/\sqrt{2}$.

Now let us analyze the spectrum of the fermion-kink system. We are interested in the fermion with a single pair of excited states, positive and negative, but with a frequency distinct from the scalar vibrational mode $\omega_S=\sqrt{3/2}\simeq 1.225$. This is important to distinguish which mode is responsible for the energy exchange mechanism and the consequent resonant structure. We take a suitable value of the coupling parameter $g=0.87$, which is not too close to the threshold value $g=1/\sqrt{2}\simeq 0.707$, where the fermion first excited state becomes non-normalizable. The spectrum is composed of the eigenvalues listed in table \ref{tab:P4-eig}, as discussed in section \ref{sec:stab}. The frequency of the excited fermionic state for this value of $g$ is $E_1^{(0)}\simeq0.855$. Moreover, there is a symmetric negative energy excitation. Both excitation frequencies can be easily distinguished from $\omega_S$.

\begin{table}
    \centering
    \begin{tabular}{c|c|c|c}
        $n$ & $\omega_n$ & $\eta_n$ & $\zeta_n$\\
        \hline
        1 & 0  & $\frac{1}{\sqrt{2}}\text{sech}^2(x/\sqrt{2})$ & $\frac{d\psi_0(x)}{dx}$ \\
        2 & 0  & 0 & $\psi_0(x)$ \\
        3 & $\sqrt{3/2}$ & $\text{sech}(x/\sqrt{2})\tanh(x/\sqrt{2})$ & No closed form\\
        4 & $E_1^{(0)}$ & 0 & $\psi_{1}^{(0)}(x)$\\
        5 & $E_{-1}^{(0)}$ & 0 & $\psi_{-1}^{(0)}(x)$
    \end{tabular}
    \caption{Solutions to the decoupled stability equation. The decoupling limit corresponds to $\alpha\to\infty$. We consider the $\phi^4$ model with $g=0.87$. The fermionic component for $n=3$ is not available in closed form.}
    \label{tab:P4-eig}
\end{table}

\subsection{Sine-Gordon model}

The sine-Gordon potential can be written as
\begin{equation}
V(\chi)=\frac{1}{2}\cos^2(\chi).
\end{equation}
For configurations where the back-reaction vanishes, the kink solution is 
\begin{equation}
\chi_k=\arcsin(\tanh(x)).
\end{equation}
Therefore, the coupling function is given by $G(\chi)=\sin(\chi)$ to have the Dirac equation in \eqref{eq_dirac_analytic}, fixing the parameter $a=1$. The function $G$ is also periodic, with the period twice as large as the potential. However, two neighboring kinks still lead to the same Dirac equation, except for a minus sign. 

In the sine-Gordon model, the kink does not possess vibrational modes, and the model is integrable, meaning that the collision output is trivial. However, with the inclusion of the fermion field, the system leads to various scattering scenarios. Fixing the coupling parameter at $g=2.0$, we obtain the discrete excitation spectrum listed in table \ref{tab:SG-eig}. It contains a positive excited fermionic state with energy $E_1^{(0)}=\sqrt{3}\simeq1.732$, and a symmetric negative energy state.

\begin{table}
    \centering
    \begin{tabular}{c|c|c|c}
        $n$ & $\omega_n$ & $\eta_n$ & $\zeta_n$\\
        \hline
        1 & 0 & $\text{sech}(x)$ & $\frac{d\psi_0(x)}{dx}$ \\
        2 & 0 & 0 & $\psi_0(x)$ \\
        3 & $E_1^{(0)}$ & 0 & $\psi_{1}^{(0)}(x)$\\
        4 & $E_{-1}^{(0)}$ & 0 & $\psi_{-1}^{(0)}(x)$
    \end{tabular}
    \caption{Solutions to the decoupled stability equation. The decoupling limit corresponds to $\alpha\to\infty$. We consider the sine-Gordon model with $g=2.0$.}
    \label{tab:SG-eig}
\end{table}

\section{Results}
\label{sec_res}

We are interested in the following collision scenario. A kink and an antikink approach each other with relative velocity $2v_i$ starting at positions $\mp X_0$, respectively. Each kink is bound to the fermion at the zero mode. Before the kinks superpose the fields are well-approximated by
\begin{align}
\chi(x,t)&=\chi_k(\gamma(x+X_0-v_it))-\chi_k(\gamma(x-X_0+v_it))-C,\\
\psi(x,t)&=\frac{1}{\sqrt{2}}\Lambda\psi_0(\gamma(x+X_0-v_it))-\frac{1}{\sqrt{2}}\Lambda^{-1}i\sigma_2\psi_0(\gamma(x-X_0+v_it)),
\end{align}
where $C$ is a constant that adjust the boundaries of the field to lie at the desired vacua. In all simulations, we chose $X_0=15.0$. The boost matrix $\Lambda$ is given by
\begin{equation}
\Lambda=\begin{pmatrix}\cosh(\nu/2)&-i\sinh(\nu/2)\\
i\sinh(\nu/2)&\cosh(\nu/2)\end{pmatrix},
\end{equation}
with the rapidity $\nu=\tanh^{-1}(v_i)$ and $i\sigma_2$ is the operator responsible for $\phi\to-\phi$ transformation, which in this case transforms the fermion zero mode in a kink background into the same state in an antikink background. This fermion configuration has nonvanishing back-reaction when the kinks superpose. For this reason, the fermion can exchange energy with the kink. On the other hand, had we started with a fermion bound only to the kink (or the antikink) at the zero mode, the back-reaction would vanish during the whole evolution of the system \cite{campos2022kink}. 

We integrate the equations of motion in an interval $-400.0<x<400.0$ discretized in steps $\Delta x=0.05$ and with periodic boundary conditions. The spatial derivatives are computed using a five-point stencil approximation and integrated using a fifth-order Runge-Kutta method with adaptive steps and error control. We measure the system's energy to test our method's accuracy. The maximum relative error during the whole evolution is of order $10^{-6}$.

\subsection{$\phi^4$ model}

\begin{figure}[tbp]
\centering
   \includegraphics[width=0.8\linewidth]{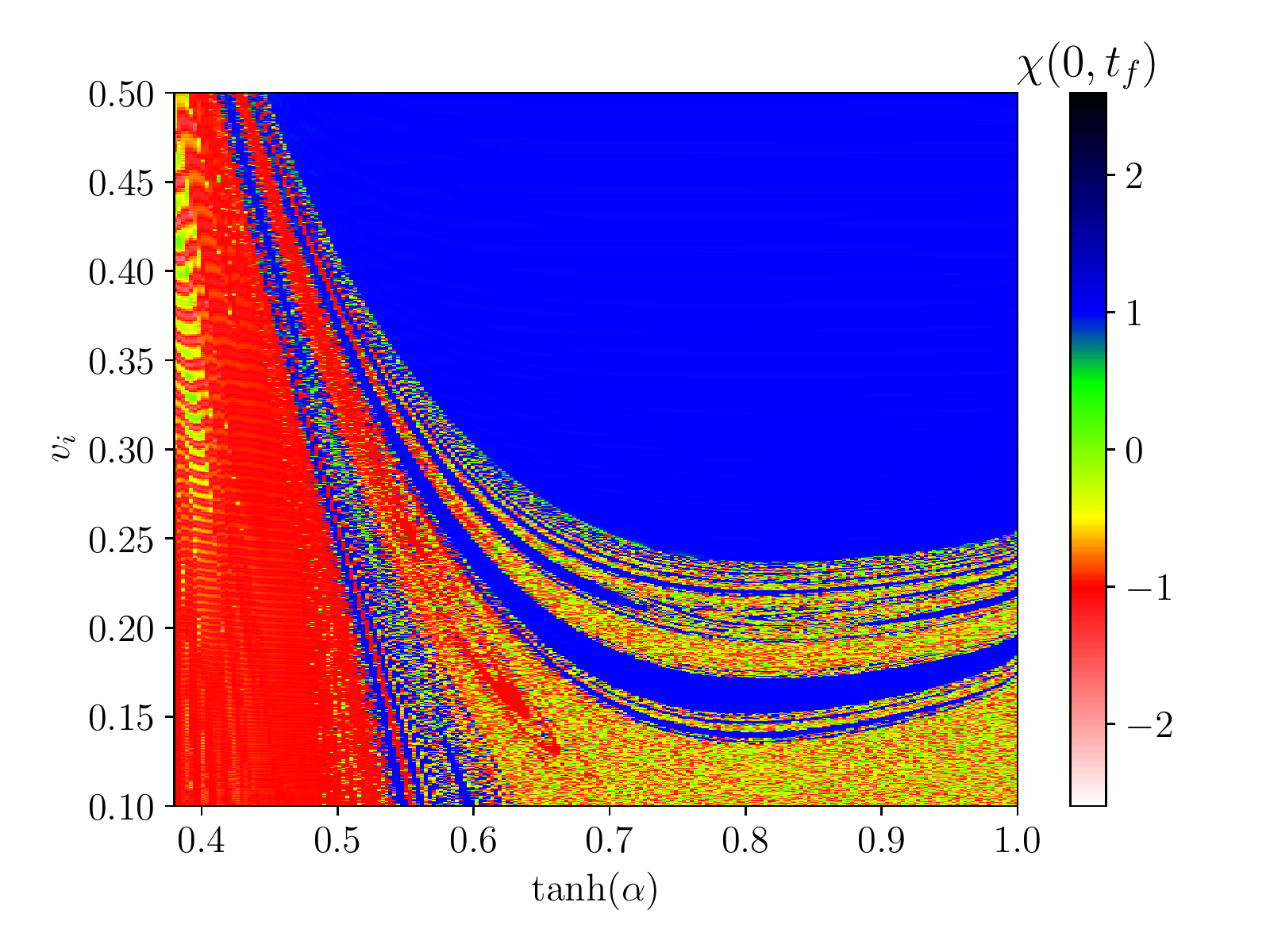}
   \caption{Final value of the field at the center of the collision as a function of \textcolor{red}{$\alpha$} and $v_i$. The final time is given by $t_f=60/v_i$. We are considering the $\phi^4$ model.}
   \label{fig_mat}
\end{figure}

 We will start focusing on the $\phi^4$ model. To access the system's behavior as a function of the parameters $v_i$ and $\alpha$, we plot the final value of the field at the center of the collision in Fig.~\ref{fig_mat}. The hyperbolic tangent is included in the horizontal axis to map large values of $\alpha$ to one. The final time is set to $t_f=60.0/v_i$. The color at every point maps to the corresponding behavior. When the kinks separate, the field at the center becomes $\chi\simeq1.0$, which is near the highest vacuum and marked blue. Therefore, the blue regions correspond to two possible behaviors. The first is a reflection after a single bounce, which can be seen in the largest blue region. The second one is resonance windows, which occur in all other regions marked in blue. The other colors correspond to kink-antikink annihilation. If a bion is formed, the field oscillates chaotically, and many different colors can appear, but mostly green and yellow. Sometimes the bion decays into two oscillons. Whenever it occurs, the field at the center reaches the value $\chi\simeq-1.0$, corresponding to the lowest vacuum and red color. As $\alpha$ is increased, the back-reaction becomes negligible, and we recover the original $\phi^4$ resonance structure. Moreover, the resonance structure is still present for low values of $\alpha$, i.e., with substantial back-reaction.

\begin{figure}[tbp]
\centering
   \includegraphics[width=\linewidth]{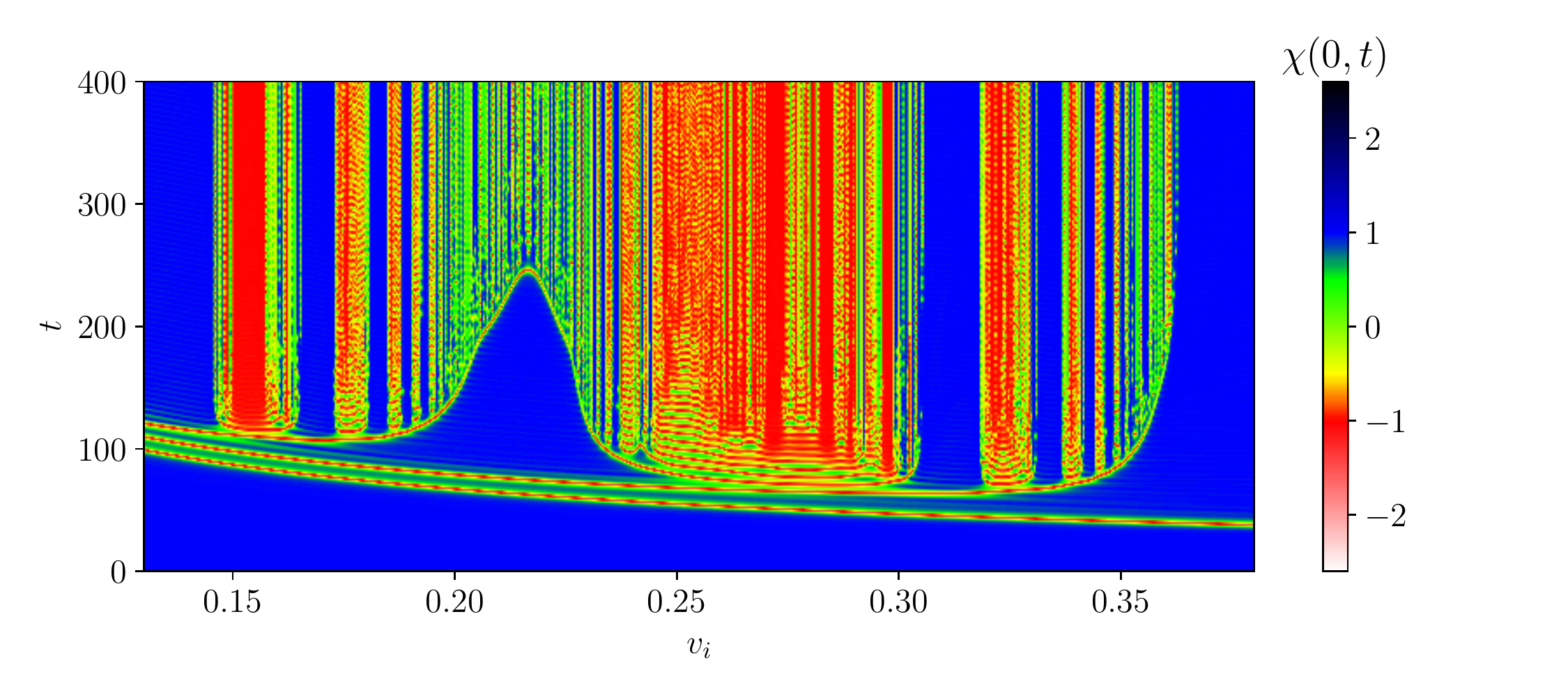}
   \caption{Field at the center of the collision as a function of $t$ and $v_i$. We consider the $\phi^4$ model and fix $\alpha=0.6$.}
   \label{fig_cent}
\end{figure}

In order to analyze the resonance structure more closely, we fix $\alpha=0.6$ and measure the field at the center of the collision as a function of time and $v_i$. The result is shown in Fig.~\ref{fig_cent}. The figure can be interpreted as follows. At every bounce, the field crosses the value $\chi=0$, which is marked by a yellow and green line. As pointed out above, the field at the center approaches $\chi\simeq1.0$ when the kinks separate, corresponding to a blue color in the figure. Therefore, blue vertical stripes depict resonance windows and reflection. The latter means separation after a single bounce. It occurs for all initial velocities greater than a critical one $v_c\simeq 0.362$. The annihilation scenario can create a bion or two oscillons. At values of $v_i$, which correspond to bion formation, the color varies as one moves vertically. As mentioned previously, the field at the center approaches $\chi\simeq-1.0$ if two oscillons are formed. Therefore, the formation of two oscillons corresponds to red vertical stripes. The figure shows that a false resonance window exists at $v_i\simeq0.215$. Moreover, there is a sequence of two bounce windows that accumulate near the critical velocity. Therefore, the resonant mechanism is clearly at play in this case.

Another property of the present model is that resonance windows with three or more bounces can be noticeably large. For instance, many three bounce windows are observed near the false resonance window, and some have considerable width. This property contrasts with the usual resonance observed in systems consisting only of scalar fields.   

\begin{figure}[tbp]
\centering
   \begin{subfigure}[b]{1.0\textwidth}         
         \centering
         \includegraphics[width=\textwidth]{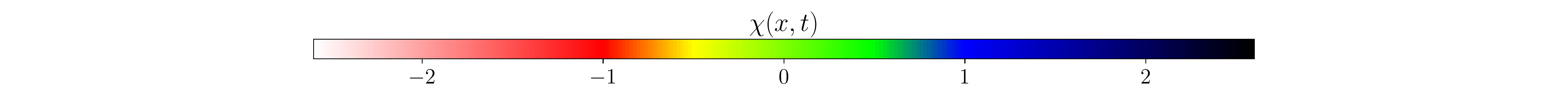}
   \end{subfigure}
     \begin{subfigure}[b]{0.32\textwidth}         
         \centering
         \includegraphics[width=\textwidth]{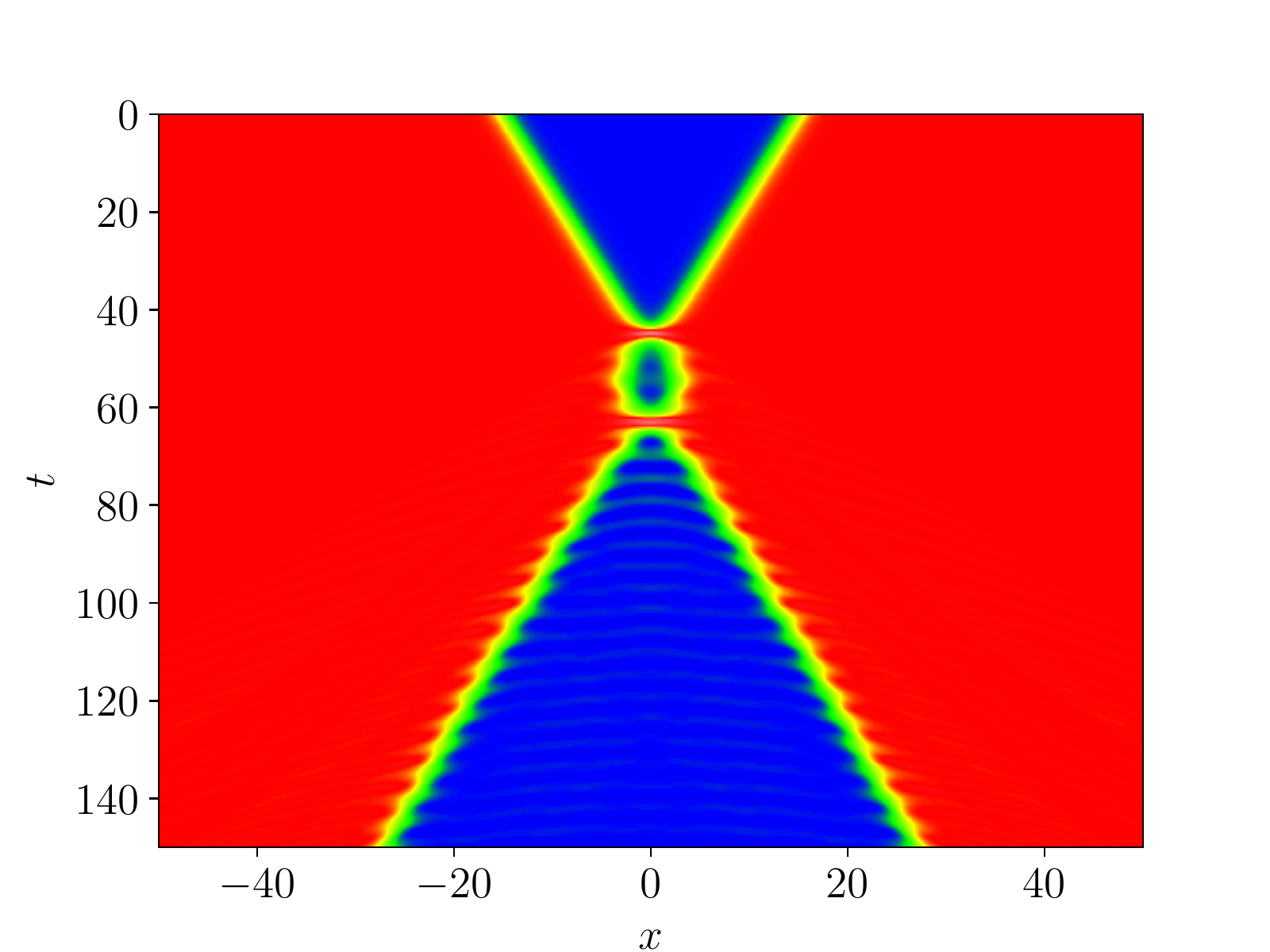}
         \caption{$\alpha=0.6$, $v_i=0.31$}
     \end{subfigure}
     \begin{subfigure}[b]{0.32\textwidth}         
         \centering
         \includegraphics[width=\textwidth]{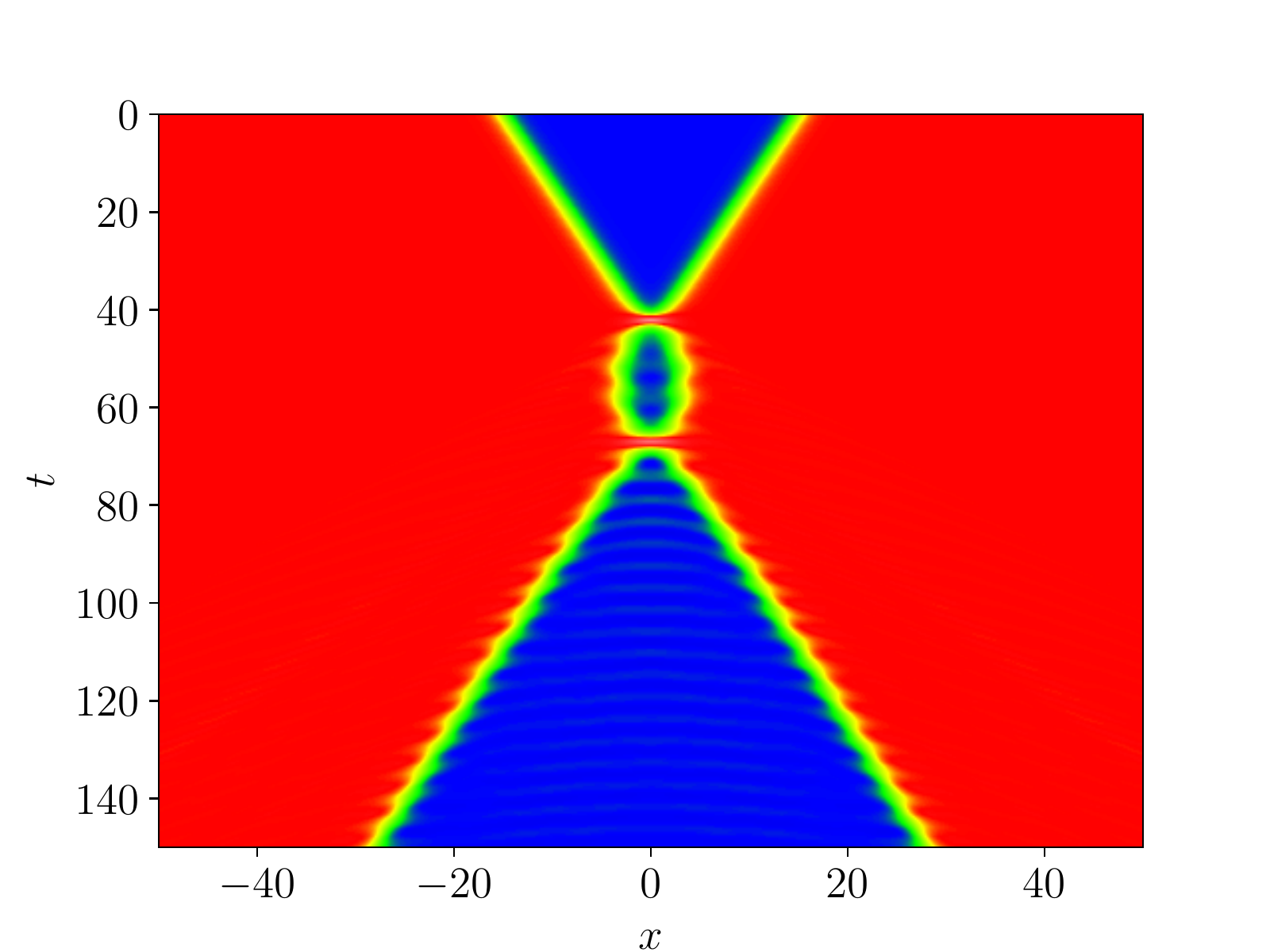}
         \caption{$\alpha=0.6$, $v_i=0.3325$}
     \end{subfigure}
     \begin{subfigure}[b]{0.32\textwidth}         
         \centering
         \includegraphics[width=\textwidth]{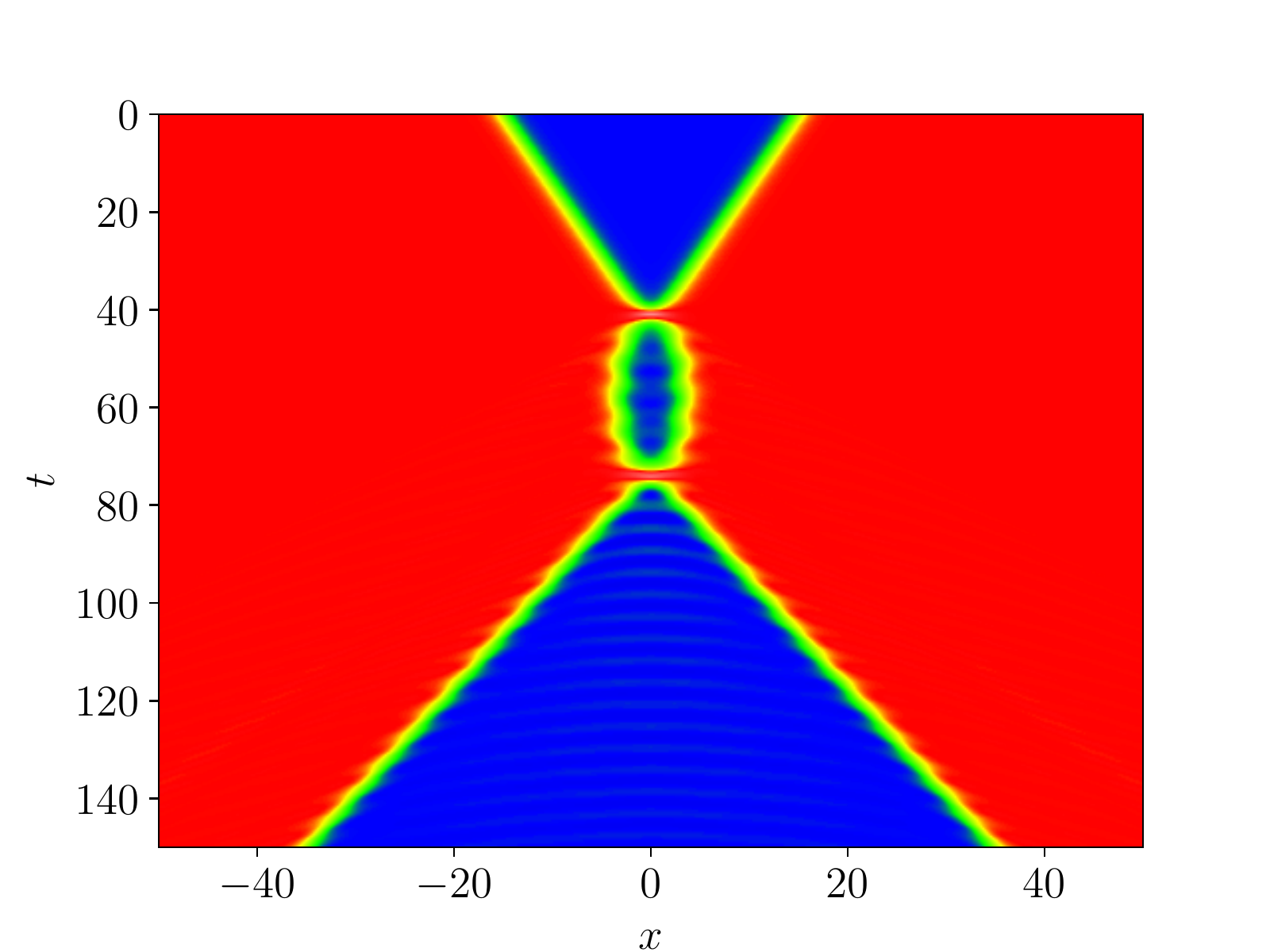}
         \caption{$\alpha=0.6$, $v_i=0.34275$}
     \end{subfigure}
     \includegraphics[width=\textwidth]{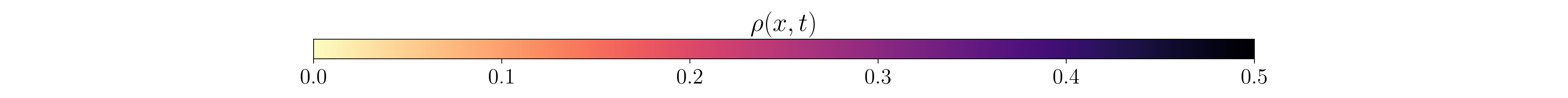}
     \begin{subfigure}[b]{0.32\textwidth}         
         \centering
         \includegraphics[width=\textwidth]{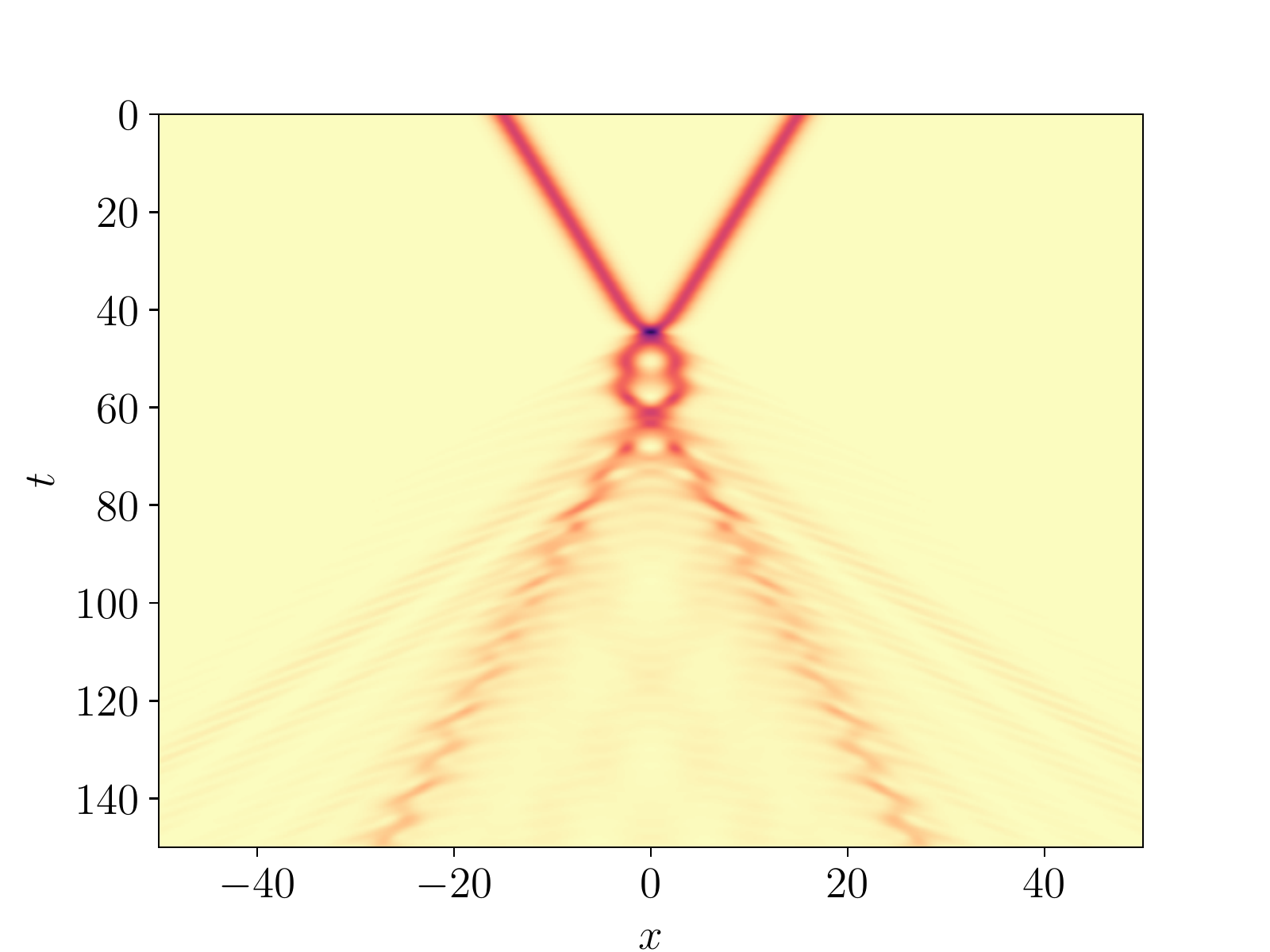}
         \caption{$\alpha=0.6$, $v_i=0.31$}
     \end{subfigure}
     \begin{subfigure}[b]{0.32\textwidth}         
         \centering
         \includegraphics[width=\textwidth]{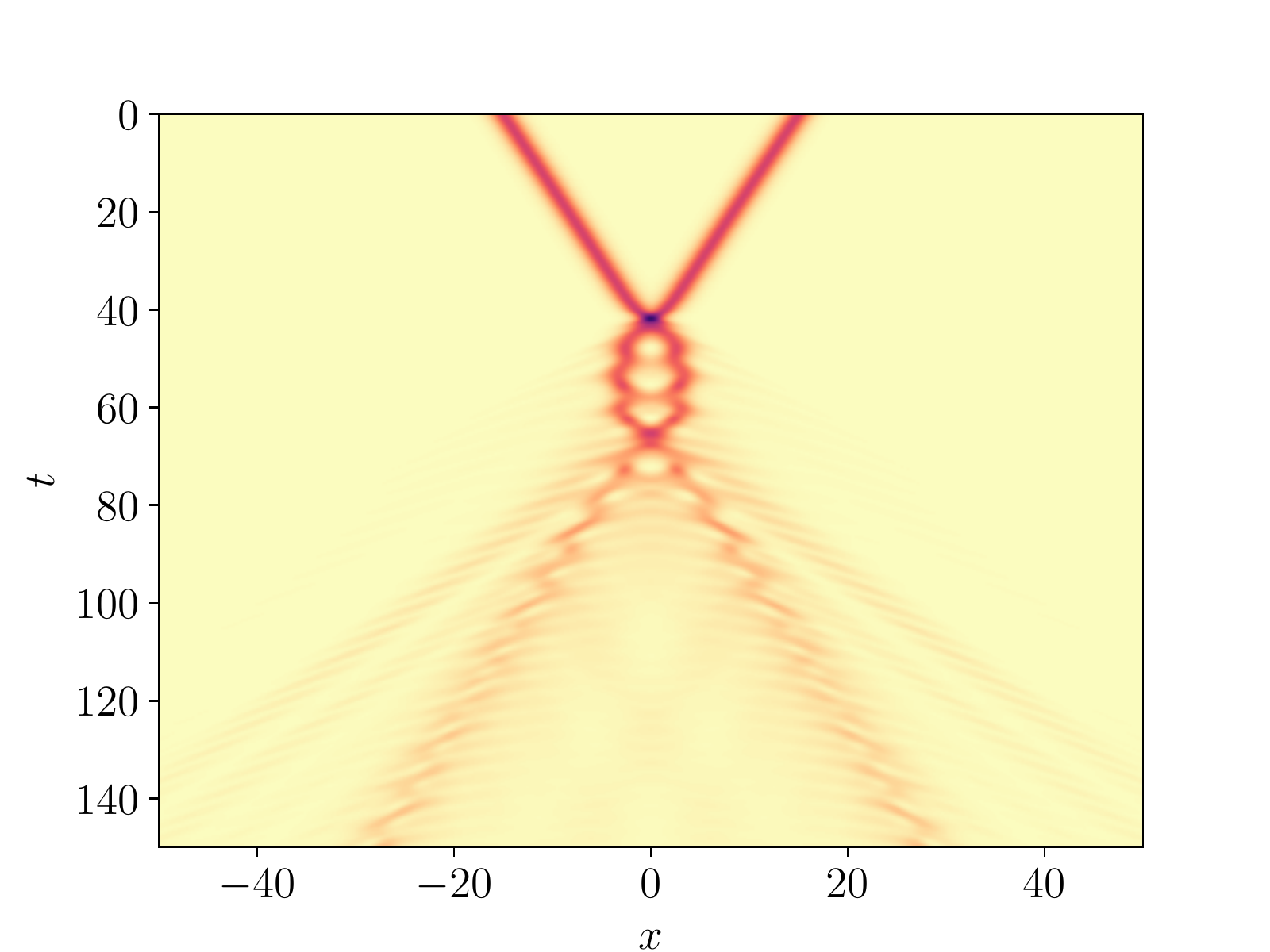}
         \caption{$\alpha=0.6$, $v_i=0.3325$}
     \end{subfigure}
     \begin{subfigure}[b]{0.32\textwidth}         
         \centering
         \includegraphics[width=\textwidth]{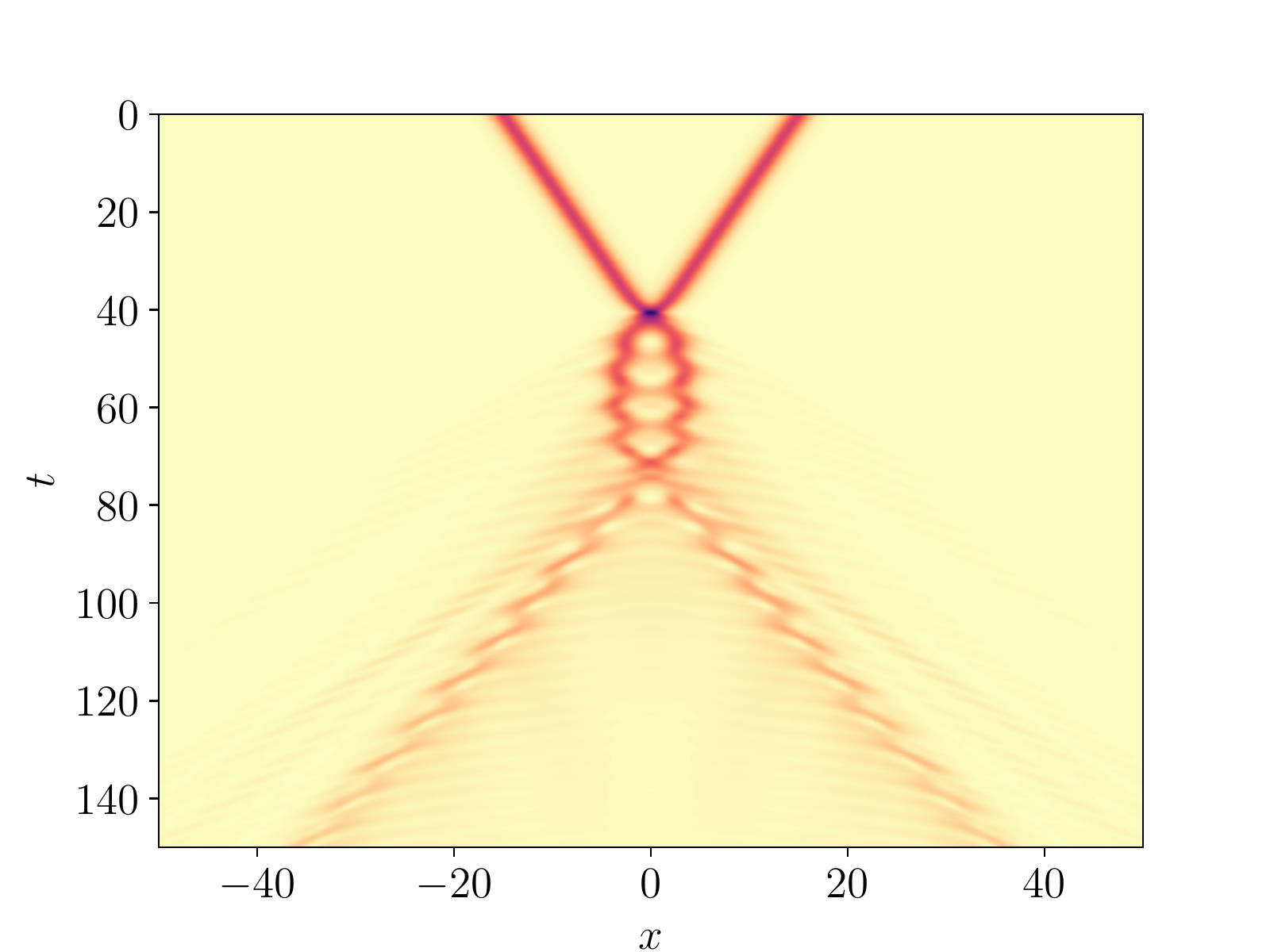}
         \caption{$\alpha=0.6$, $v_i=0.34275$}
     \end{subfigure}
     \caption{Evolution of the scalar field (upper) and fermion density (lower) in spacetime. We consider the first three two-bounce windows of the $\phi^4$ model with $\alpha=0.6$.}
   \label{fig_fermion}
\end{figure}

We examine the model in the first three resonance windows in Fig.~\ref{fig_fermion}. We plot both the scalar field and the fermion density $\rho(x,t)\equiv\psi^\dagger(x,t)\psi(x,t)$. The main element that should be observed is that the fermion exhibits a well-defined oscillation between the first and second bounces. Furthermore, the number of oscillations increases by one from one window to the next. Similar behavior is observed for the scalar field, although slightly less pronounced. This suggests that some vibrational mode is responsible for resonant behavior.

\begin{figure}[tbp]
\centering
   \includegraphics[width=0.6\linewidth]{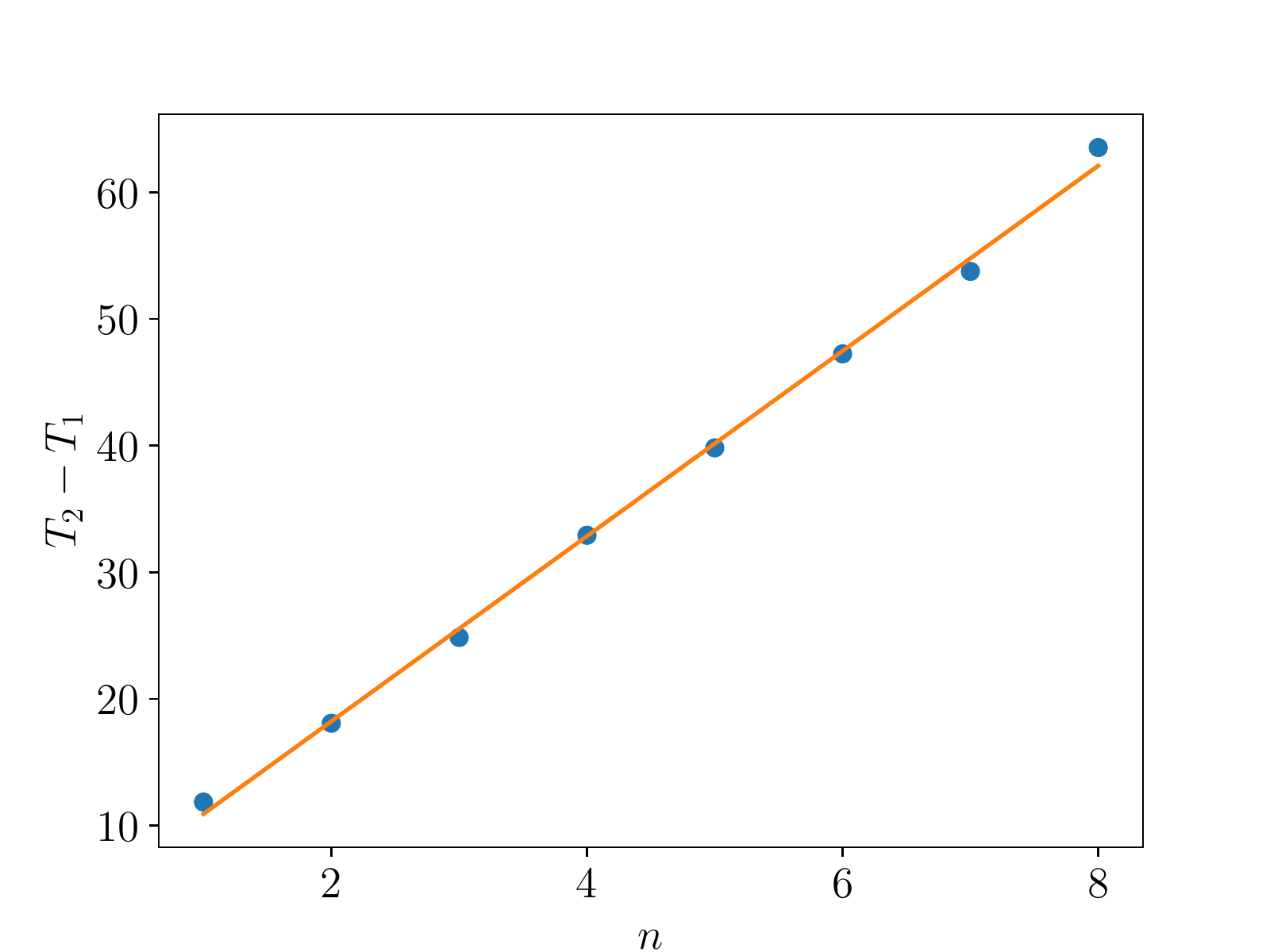}
   \caption{Time between the first and second bounces as a function of the window number. We consider the $\phi^4$ model and fix $\alpha=0.6$. The best fitted straight line is also shown.}
   \label{fig_times}
\end{figure}

In order to tell which mode is responsible for the energy exchange mechanism, we need to measure the vibrational mode's frequency $\omega_R$. It can be obtained by measuring the time between the first and second bounces $T\equiv T_2-T_1$. We plot this quantity as a function of the window number $n$. The false resonance window corresponds to $n=1$, and we increase it by one as we move to the next window. The result is shown in Fig.~\ref{fig_times}. As shown in Ref.~\cite{campbell1983resonance}, the linear behavior is expected from the relation
\begin{equation}
\label{eq_camp}
\omega_R T=\delta+2\pi n,
\end{equation}
where $\delta$ is a constant shift. Performing a numerical fit, we find $\omega_R=0.859$. Clearly, the exchange mechanism does not come from the kink's vibrational mode, which corresponds to the frequency $\omega_S\simeq1.225$. When compared to the fermion excited state's frequency, the relative error is
\begin{equation}
e=\frac{|\omega_R-E_1^{(0)}|}{E_1^{(0)}}\simeq 0.57\%.
\end{equation}
This result is highly assuring that the fermion is mediating the exchange mechanism.

\subsection{Sine-Gordon model}
To provide more decisive proof, let us proceed to the sine-Gordon model, which is integrable in the absence of the fermion field. It means that the kink-antikink collision is elastic in this case. The sine-Gordon model does not have any vibrational mode. Therefore, if we can observe a resonance structure when the kink is coupled with a fermion field, the only possible explanation for the energy exchange mechanism could originate from the presence of the fermion field. With the analysis made in sec. 2.3, it becomes easy to associate the resonance structure to the fermion excitation frequency.

\begin{figure}[tbp]
\centering
   \includegraphics[width=0.8\linewidth]{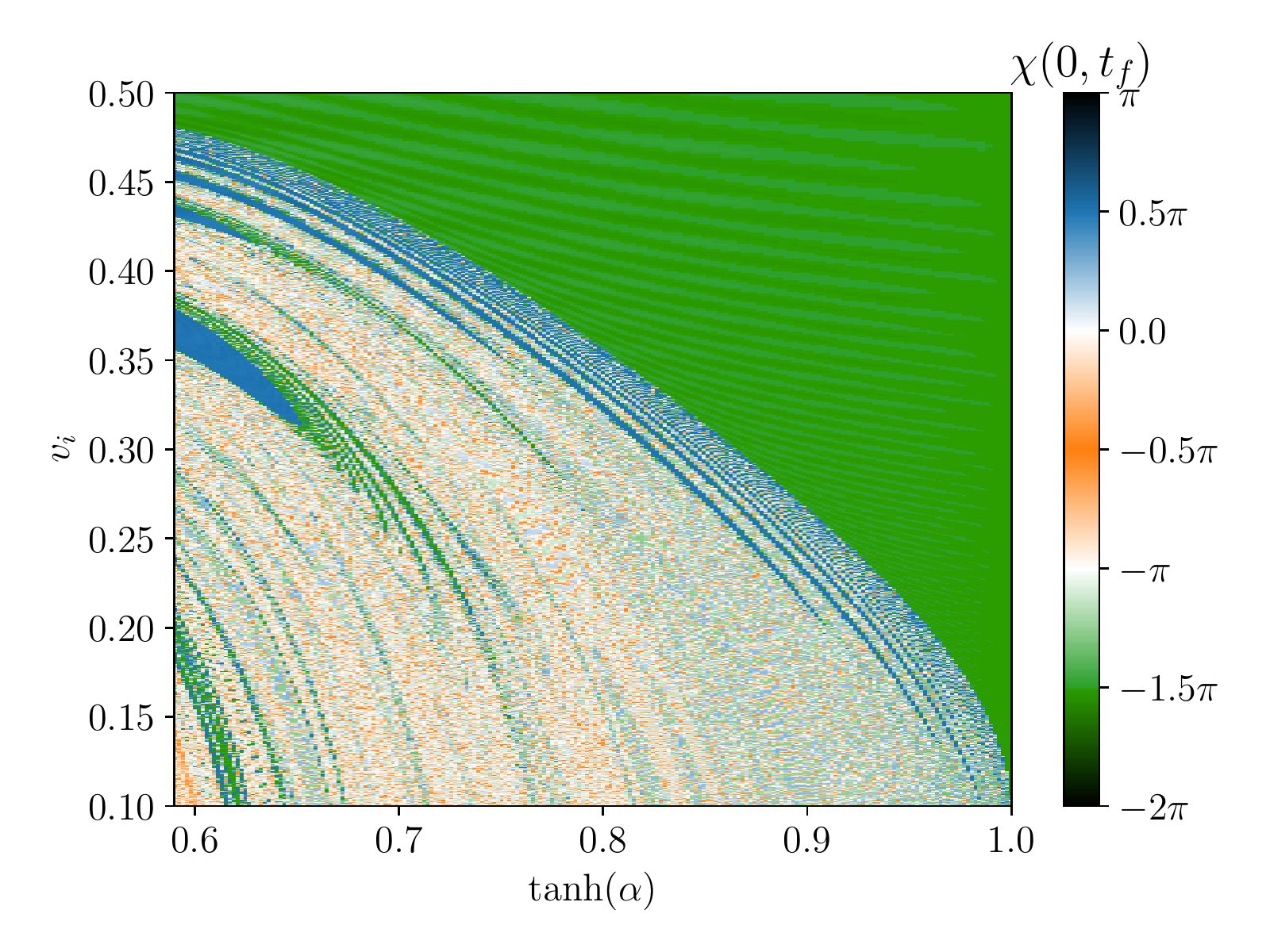}
   \caption{Final value of the field at the center of the collision as a function of $\alpha$ and $v_i$. The final time is given by $t_f=60/v_i$. We are considering the sine-Gordon model.}
   \label{fig_mat_SG}
\end{figure}

The system's behavior as a function of the parameters $v_i$ and $\alpha$ is summarized in Fig.~\ref{fig_mat_SG}. Again, we include a hyperbolic tangent to map large values of $\alpha$ to one. It shows the value of the scalar field at the center of the collision at a final time $t_f=60.0/v_i$. The figure can be read as follows. If the kinks annihilate, they form a bion state, which oscillates around the vacuum $-\pi/2$, marked in orange. In this region, $\chi(0,t_f)$ can have many different colors, but mainly orange and white. If the kinks separate, it can occur after an odd (even) number of bounces, which means that they cross (reflect). This is possible because the scalar potential is periodic, and the kinks can change sectors. They leave a vacuum $\chi=-1.5\pi$ at the center when they cross.
In the figure, the corresponding region is marked in green. If they reflect, the center is at the vacuum $\chi=0.5\pi$ and is marked blue. This figure has many interesting features. First, the large green region corresponds to crossing after a single bounce. The value of $v_i$ that separates this region gives the critical velocity. It is possible to see a sequence of two bounce windows accumulating near the critical velocity for all values of \textcolor{red}{$\alpha$} that we considered.
Moreover, the critical velocity decreases with $\alpha$. The decrease becomes very sharp as $\tanh(\alpha)$ approaches one. Such a result is expected because the original sine-Gordon model is integrable and, as a result, has a vanishing critical velocity when $\alpha$ goes to infinity. Finally, it is possible to observe many lower-order resonance windows being created as $\alpha$ is decreased. This is a consequence of increasing the back-reaction.

\begin{figure}[tbp]
\centering
   \includegraphics[width=1.0\linewidth]{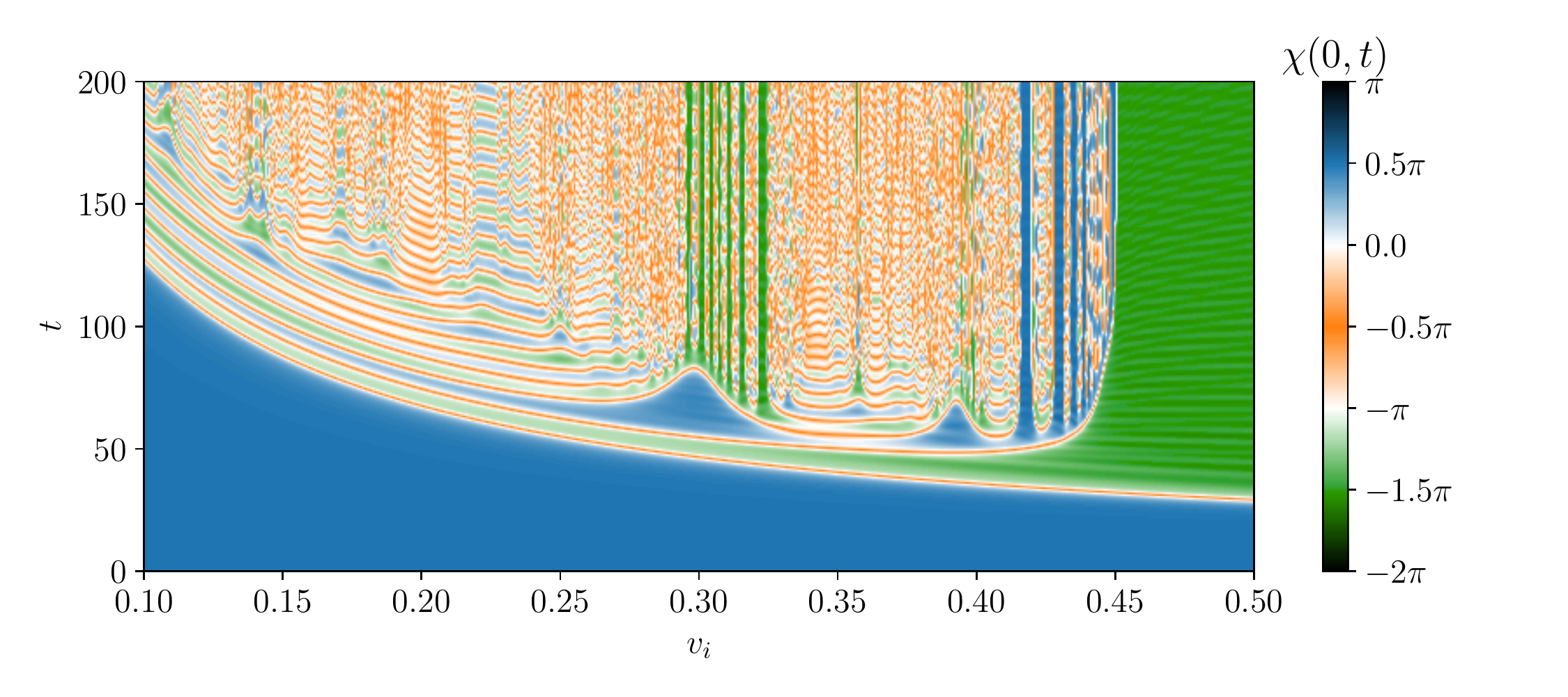}
   \caption{Field at the center of the collision as a function of $t$ and $v_i$. We consider the sine-Gordon model and fix $\alpha=0.8$.}
   \label{fig_cent_SG}
\end{figure}

Let us analyze the resonant behavior of the system in detail for a fixed value $\alpha=0.8$. The collision scenarios for a range of $v_i$ are summarized in Fig.~\ref{fig_cent_SG}. We plot the scalar field at the collision center as a function of $t$ and $v_i$. A bounce occurs when the field at the center crosses the value $\chi=-0.5\pi$, marked in orange. Three scenarios are observed. The first one corresponds to kink-antikink annihilation and formation of a bion. As a result, many orange lines are observed as we move vertically. The second case consists of kink-antikink crossing after an odd number of bounces, which leaves a vacuum $\chi=-1.5\pi$ at the center. In the figure, the corresponding values of $v_i$ exhibit green vertical stripes. If there is separation after an even number of bounces, the center is at the vacuum $\chi=0.5\pi$ leading to blue vertical stripes. In the figure, we observe two false resonance windows at $v_i\simeq 0.30$ and $v_i\simeq 0.39$. In addition, there appears a sequence of two-bounce resonance windows which accumulate near the critical velocity. 

\begin{figure}[tbp]
\centering
   \begin{subfigure}[b]{1.0\textwidth}         
         \centering
         \includegraphics[width=\textwidth]{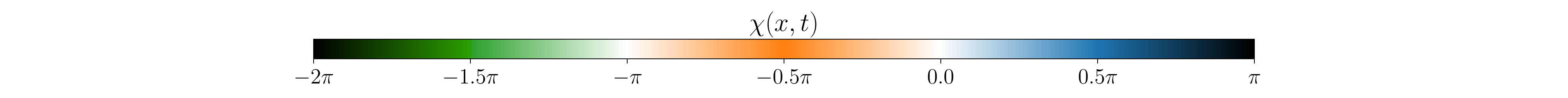}
   \end{subfigure}
     \begin{subfigure}[b]{0.32\textwidth}         
         \centering
         \includegraphics[width=\textwidth]{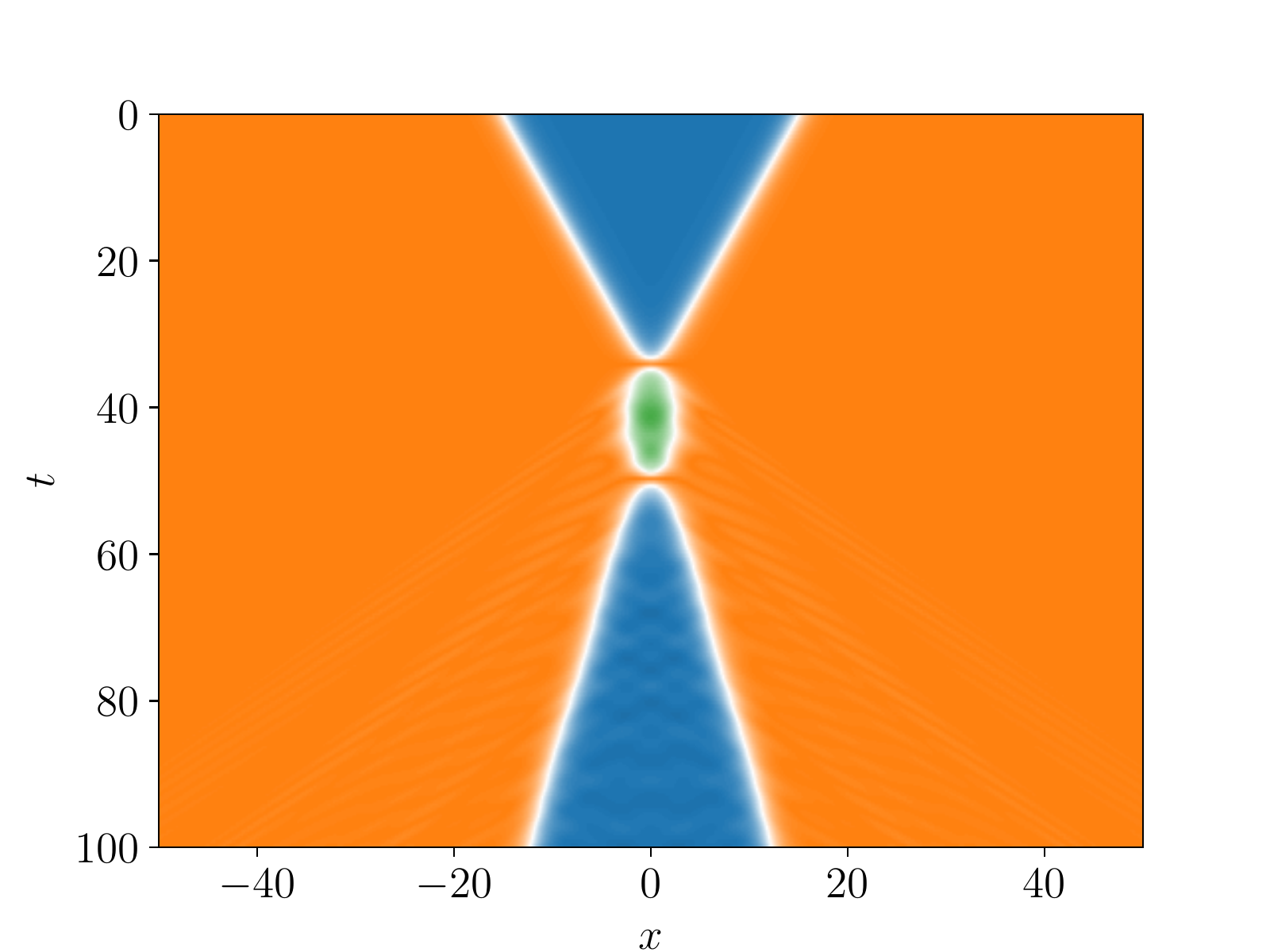}
         \caption{$\alpha=0.8$, $v_i=0.4176$}
     \end{subfigure}
     \begin{subfigure}[b]{0.32\textwidth}         
         \centering
         \includegraphics[width=\textwidth]{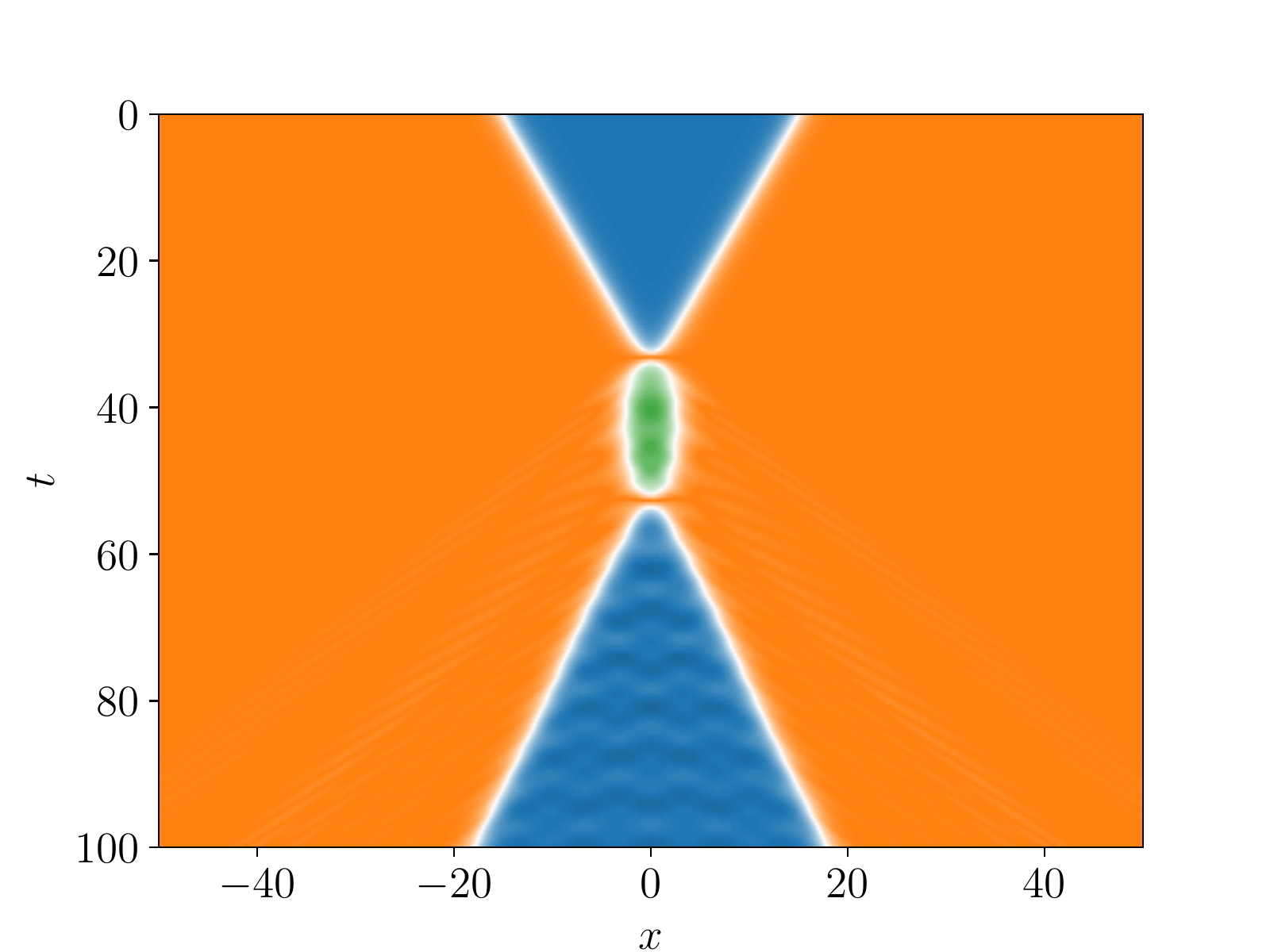}
         \caption{$\alpha=0.8$, $v_i=0.4294$}
     \end{subfigure}
     \begin{subfigure}[b]{0.32\textwidth}         
         \centering
         \includegraphics[width=\textwidth]{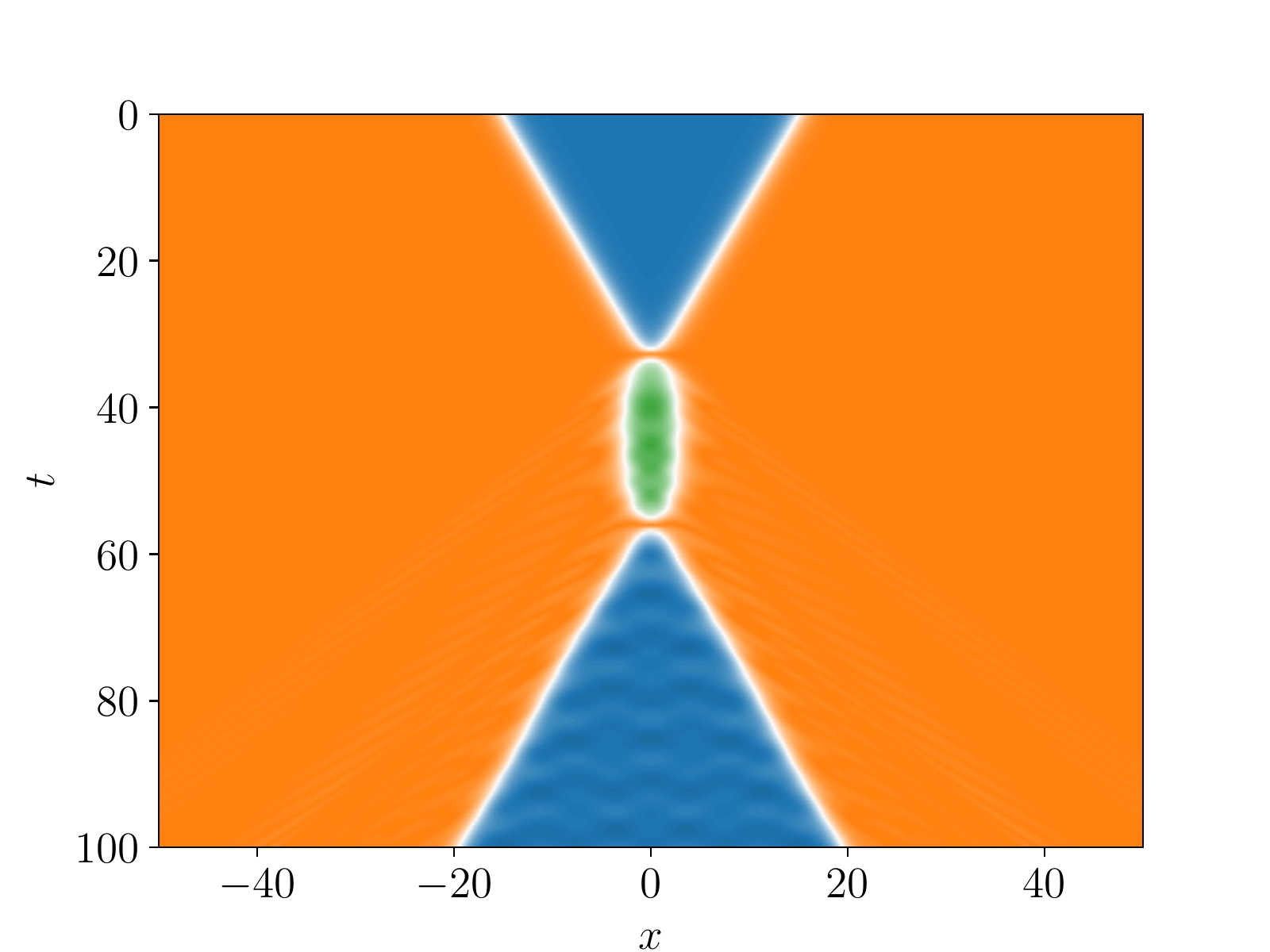}
         \caption{$\alpha=0.8$, $v_i=0.4348$}
     \end{subfigure}
     \includegraphics[width=\textwidth]{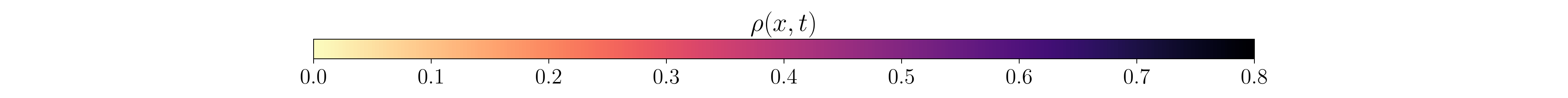}
     \begin{subfigure}[b]{0.32\textwidth}         
         \centering
         \includegraphics[width=\textwidth]{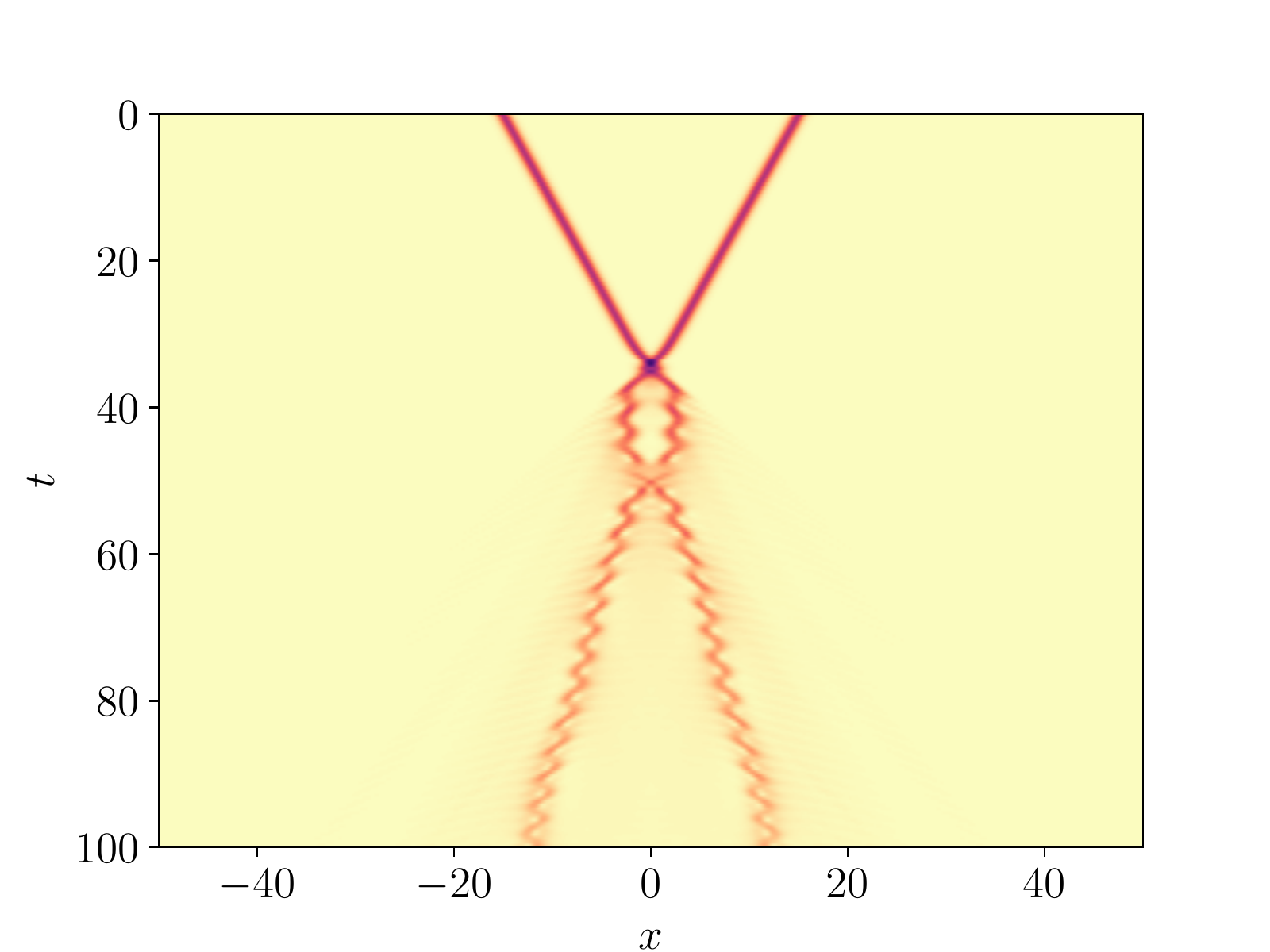}
         \caption{$\alpha=0.8$, $v_i=0.4176$}
     \end{subfigure}
     \begin{subfigure}[b]{0.32\textwidth}         
         \centering
         \includegraphics[width=\textwidth]{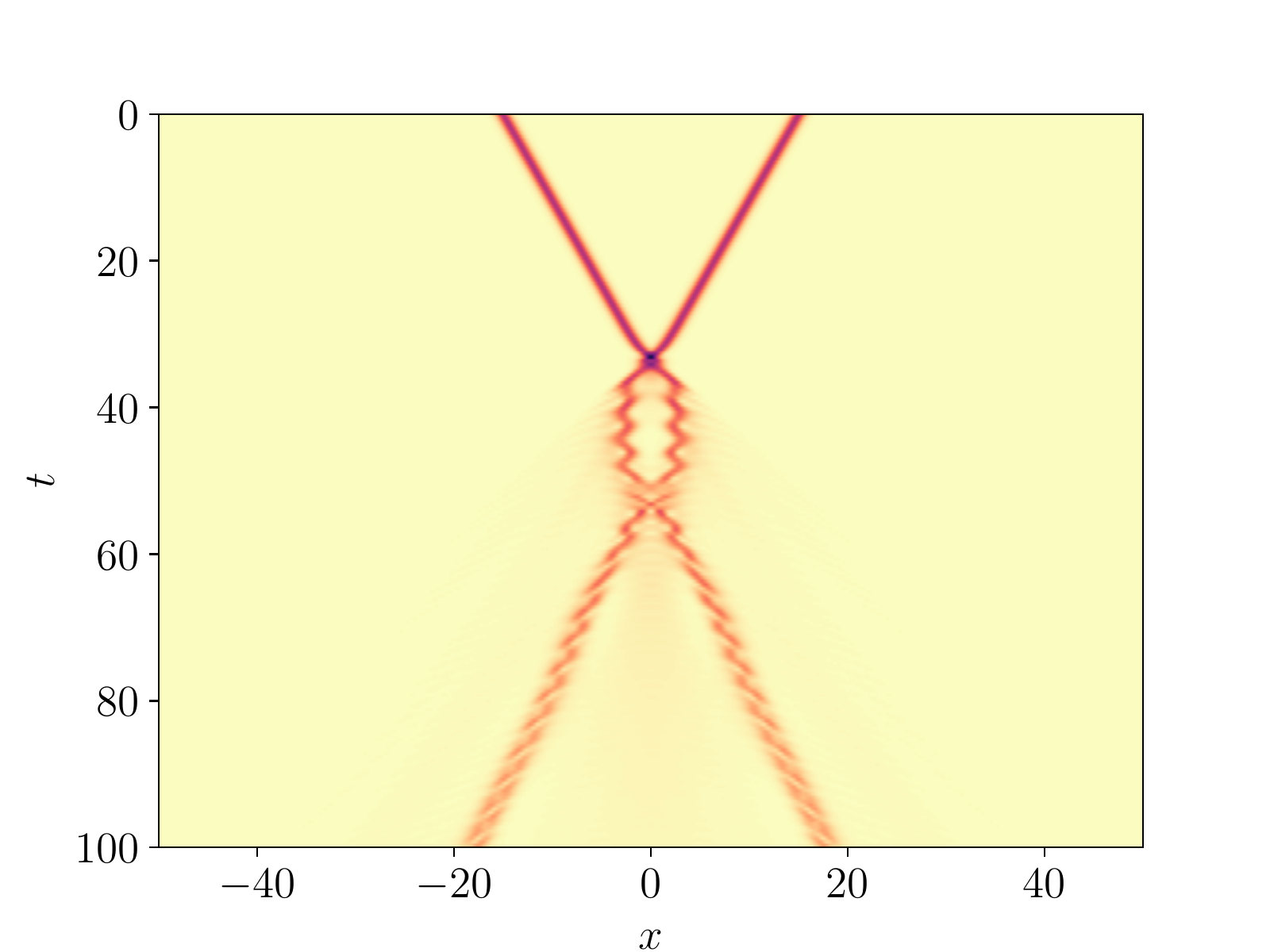}
         \caption{$\alpha=0.8$, $v_i=0.4294$}
     \end{subfigure}
     \begin{subfigure}[b]{0.32\textwidth}         
         \centering
         \includegraphics[width=\textwidth]{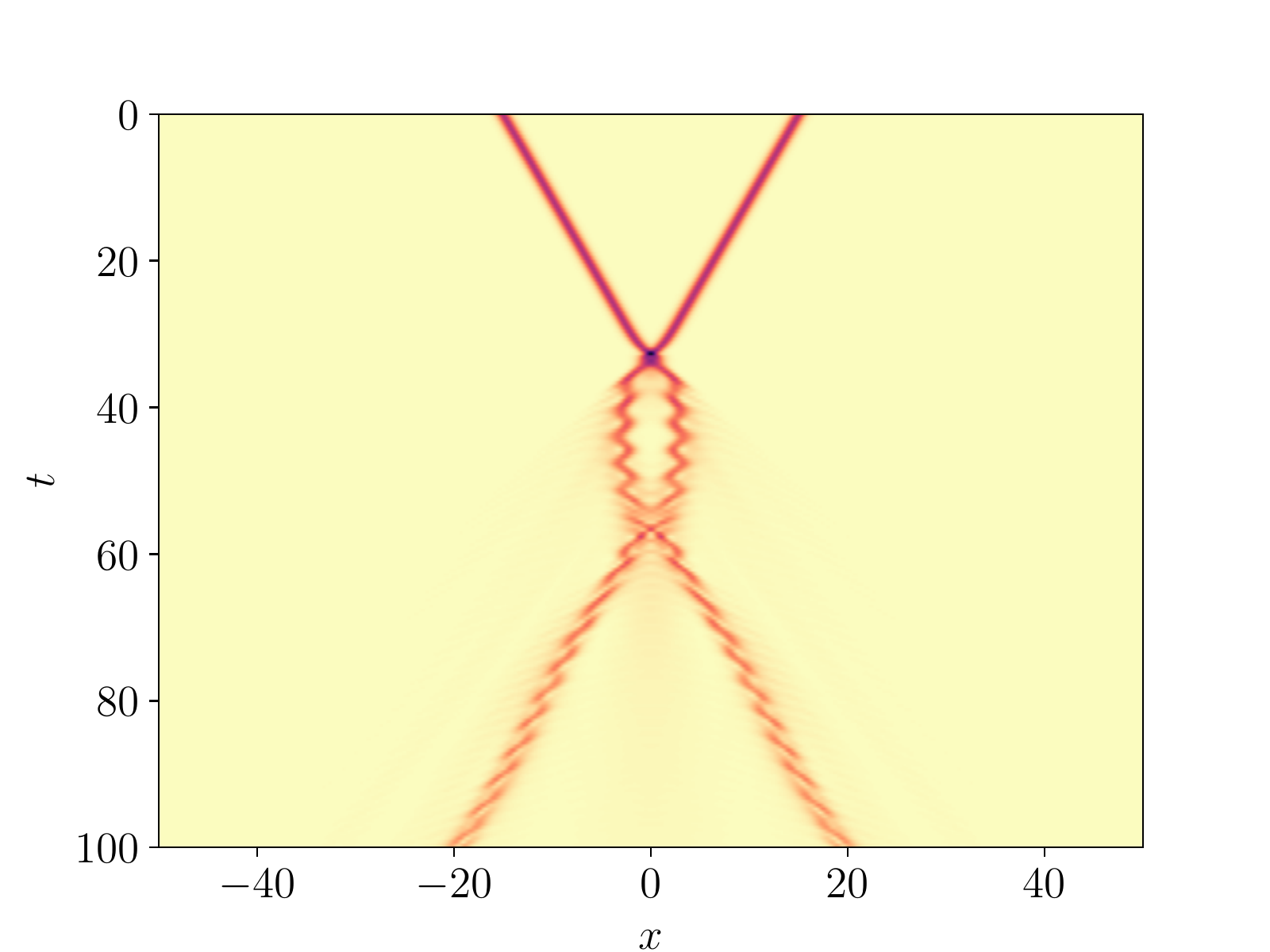}
         \caption{$\alpha=0.8$, $v_i=0.4348$}
     \end{subfigure}
     \caption{Evolution of the scalar field (upper) and fermion density (lower) in spacetime. We consider the first three two bounce windows of the sine-Gordon model with $\alpha=0.8$.}
   \label{fig_fermion_SG}
\end{figure}

Considering the appearance of resonance windows, a resonant mechanism should be at play here. To investigate further, we computed the evolution of the scalar field and fermion density in spacetime. The result is shown in Fig.~\ref{fig_fermion_SG} for the three first resonance windows. Between the first two bounces, one can see well-defined oscillation in the fermion densities. Moreover, the number of oscillations increases by one if we increase the window number by the same amount. The same pattern can be observed in the scalar field near the center but with a smaller amplitude. Therefore, it can be concluded that there is some vibrational mode responsible for the resonant behavior. 
The frequency of the mode mentioned above can be found by measuring the time between the first and second bounces $T\equiv T_2-T_1$. When plotted as a function of the window number, we should find the relation given by eq.~(\ref{eq_camp}).
The slope of the straight line fit gives the frequency $\omega_R\simeq1.671$. The corresponding error concerning the fermion excited state energy is
\begin{equation}
e=\frac{|\omega_R-E_1^{(0)}|}{E_1^{(0)}}\simeq 3.54\%.
\end{equation}
The agreement is still notable, considering all approximations in the analysis. Therefore, we can conclude that it is the fermion excited state that is mediating the exchange mechanism.
\begin{figure}[tbp]
\centering
   \includegraphics[width=0.6\linewidth]{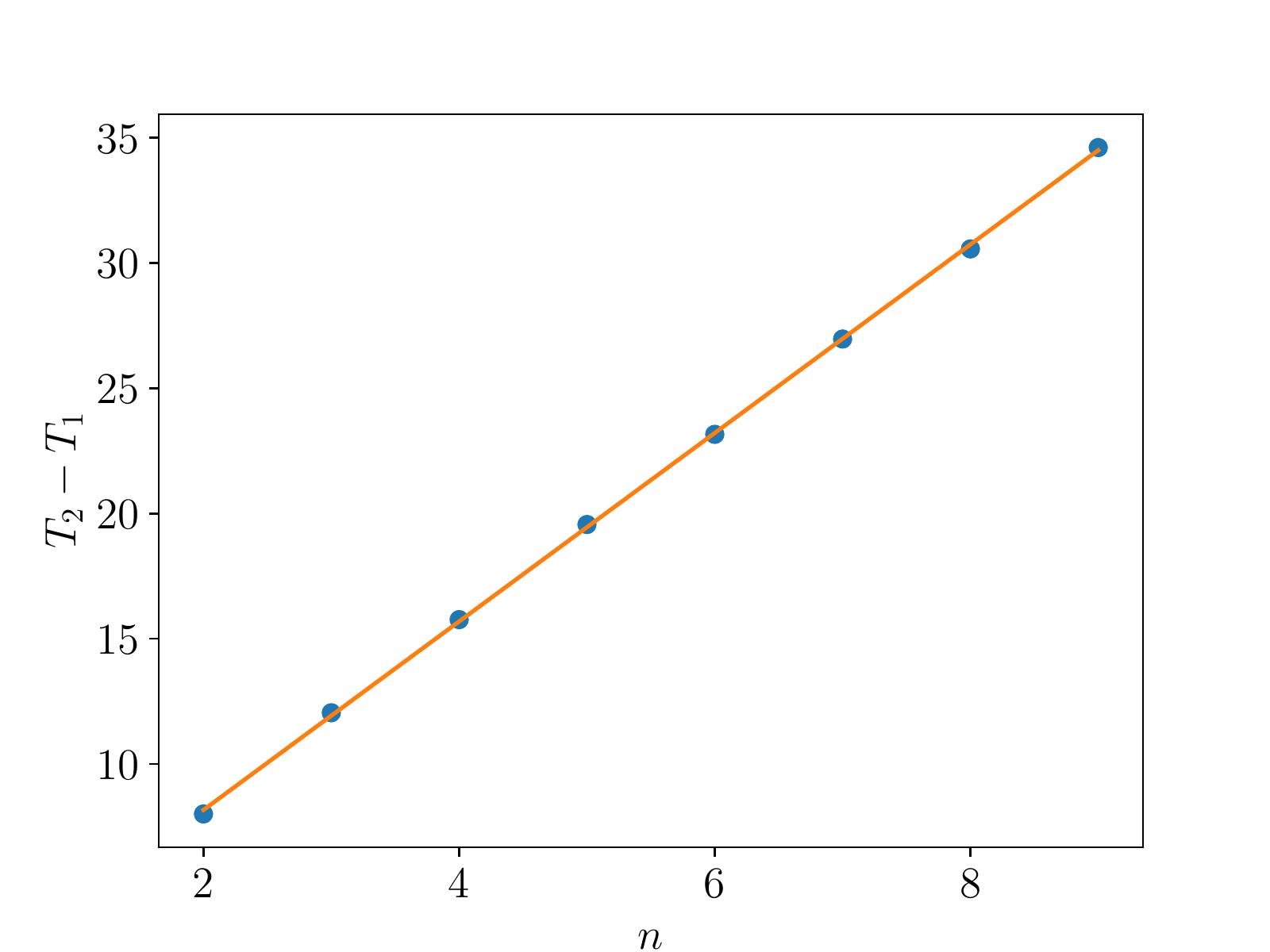}
   \caption{Time between the first and second bounces as a function of the window number. We consider the sine-Gordon model and fix $\alpha=0.8$. The best fitted straight line is also shown.}
   \label{fig_times_SG}
\end{figure}

\section{Conclusion}
\label{sec_conc}

In this work, we analyzed two models that describe the interaction between fermions and kinks. They are extensions of the $\phi^4$ and sine-Gordon models. The coupling between the fermion and scalar fields was chosen to obtain a P\"oschl-Teller form for the effective fermionic potential in the absence of back-reaction. The usual Yukawa potential is considered for the $\phi^4$ model, while we take a periodic interaction for the sine-Gordon one. 

The parameters were chosen to fix the number of fermion excited states to a single pair containing a positive energy state and a negative one. The $\phi^4$ model fermion excitation frequency could be easily distinguished from the scalar field vibrational one because they are considerably apart. The sine-Gordon creates a different scenario where the kink possesses no vibrational mode. Thus, the possibility of storing energy in a vibrational mode comes from including a fermion field.

Considering kink-antikink collisions where both kinks start with fermions at the zero mode, we obtained a rich resonant structure in both models, varying the strength of the back-reaction. We focused on a single case with strong back-reaction to comprehend better the resonant structure. In resonance windows, we observed oscillations in the fermion density between bounces which increase by one as the window number is increased by the same amount. The frequency of such oscillation was measured, and the obtained values were close to the theoretical energy of the first fermion excited state. 

Our work shows that adding a fermion field can lead to the formation of resonance windows. Surprisingly, the observed exchange mechanism is mediated by the fermion excited states. Due to the coupling between the fields, this vibrational mode generates oscillations in both scalar and fermionic fields. Still, the origin of such modes comes from the inclusion of the fermion. This is the first observation of this phenomenon and provides one more exception to the usual resonant energy exchange mechanism of kink-antikink collisions.

There are many possible extensions of our work with kink-antikink collisions including the fermion back-reaction, which have not been explored well so far. One interesting possibility is to look for systems with fermionic potentials which have excited state only for a kink-antikink pair.  Moreover, the consequences of our work are numerous. It opens new possibilities for observing new phenomena due to the interaction between topological and non-topological localized structures with a fermion field, both theoretically and experimentally. For example,
 in \cite{vanhaverbeke2008control}, the presence of current pulses were used to control the domain wall polarity in a magnetic nanowire. More generally, we expect that in most scenarios, an interacting fermion will have noticeable effects on localized structures' static and dynamical properties. 

\section*{Acknowledgments}

We acknowledge financial support from the National Council for Scientific and Technological Development - CNPq, Grant No. 303469/2019-6 (DB), No. 150166/2022-2 (JGFC) and No. 309368/2020-0 (AM). DB thanks Paraíba State Research Foundation, FAPESQ-PB, Grant N. 0015/2019. AM thanks financial support from the Brazilian agency CAPES and from Universidade Federal de Pernambuco Edital Qualis A. The simulations presented here were conducted in the SDumont cluster from the Brazilian laboratory LNCC (Laborat\'orio Nacional de Computa\c{c}\~ao Cient\'ifica).

\end{document}